%% file: 00_main.tex
\documentclass[a4paper,11pt]{article}
\usepackage[a4paper, total={6in, 9in}]{geometry}
\usepackage{xspace}
\newcommand{\q}[1]{``#1''}

\newcommand*\anue{\ensuremath{\overline{\nu}_e}\xspace}
\newcommand*\inversebetadecay{\ensuremath{\anue+\mathrm{p}\longrightarrow \mathrm{e}^++\mathrm{n}}\xspace}

\usepackage[font=footnotesize, labelfont=bf,
            format=plain, labelformat=simple, 
            labelsep=endash, justification=centerlast,
            singlelinecheck=on]{caption} 
\usepackage{subcaption} 

\usepackage[
    backend=bibtex,
    bibencoding=utf8,
    style=numeric-comp,
    autolang=other,       
    sorting=none,
    doi=false, isbn=false, eprint=false, url=false,
    giveninits=true, maxnames=2,
    date=year
]{biblatex}
\renewbibmacro{in:}{}
\usepackage{authblk}
\usepackage[svgnames,table]{xcolor}
\usepackage{bm}
\usepackage{float}
\usepackage{physics}
\usepackage{mathtools}
\usepackage[load=accepted,range-phrase=--,number-unit-product=\text{ },range-units=single]{siunitx}
\usepackage{subfiles} 

\usepackage{multirow}
\usepackage{booktabs}
\usepackage{amssymb}
\usepackage{pifont}
\usepackage{lineno}
\usepackage{lineno-ams}
\usepackage{pdfcomment}
\usepackage{booktabs}
\usepackage{adjustbox}
\pdfcommentsetup{color={1.0 0.9 0.5}, open=true}

\usepackage[inline]{enumitem}
\newlist{parlist}{enumerate}{1}
\setlist[parlist,1]{label=\arabic*),itemjoin={\\},wide=\parindent,nosep,noitemsep}

\title{Vertex and Energy Reconstruction in JUNO with Machine Learning Methods}
\author[a]{Zhen~Qian}
\author[b]{Vladislav~Belavin}
\author[c,d]{Vasily~Bokov}
\author[e]{Riccardo~Brugnera}
\author[e]{Alessandro~Compagnucci}
\author[b,c]{Arsenii~Gavrikov}
\author[e]{Alberto~Garfagnini}
\author[c]{Maxim~Gonchar}
\author[b]{Leyla~Khatbullina}
\author[a]{Ziyuan~Li\footnote{Corresponding author: liziyuan3@mail.sysu.edu.cn}}
\author[f]{Wuming~Luo}
\author[c]{Yury~Malyshkin\footnote{Corresponding author: yum@jinr.ru}}
\author[e]{Samuele~Piccinelli}
\author[g]{Ivan~Provilkov}
\author[b]{Fedor~Ratnikov}
\author[c]{Dmitry~Selivanov}
\author[c]{Konstantin~Treskov}
\author[b]{Andrey~Ustyuzhanin}
\author[e]{Francesco~Vidaich}
\author[a]{Zhengyun~You}
\author[a]{Yumei~Zhang}
\author[a]{Jiang~Zhu}
\author[e]{Francesco~Manzali}

\affil[a]{Sun Yat-sen University, Guangzhou, China}
\affil[b]{National Research University Higher School of Economics, Moscow, Russia}
\affil[c]{Joint Institute for Nuclear Research, Dubna, Russia}
\affil[d]{Physics Department of Lomonosov Moscow State University, Moscow, Russia}
\affil[e]{Physics and Astronomy Department, Padova University and INFN Padova, Padova, Italy}
\affil[f]{Institute for High Energy Physics, Beijing, China}
\affil[g]{Moscow Institute of Physics and Technology, Dolgoprudny, Russia}

\begin{document}
\maketitle
\begin{abstract}
The Jiangmen Underground Neutrino Observatory (JUNO) is an experiment designed to
study neutrino oscillations. Determination of neutrino mass ordering and precise measurement of
neutrino oscillation parameters $\sin^2 2\theta_{12}$, $\Delta m^2_{21}$ and $\Delta m^2_{32}$ are the main
goals of the experiment. A rich physical program beyond the oscillation analysis is also foreseen.
The ability to accurately reconstruct particle interaction events in JUNO is of great importance for
the success of the experiment.

In this work we present a few machine learning approaches applied to the vertex and the energy
reconstruction. Multiple models and architectures were compared and studied, including
Boosted Decision Trees (BDT), Deep Neural Networks (DNN), a few kinds of
Convolution Neural Networks (CNN), based on ResNet and VGG, and a Graph Neural Network based on DeepSphere.

Based on a study, carried out using the dataset, generated by the official JUNO software,
we demonstrate that
machine learning approaches achieve the necessary level of accuracy for reaching
the physical goals of JUNO: $\sigma_E=3\%$ at $E_\text{vis}=\SI{1}{\MeV}$ for the energy and
$\sigma_{x,y,z}=\SI{10}{\cm}$ at $E_{\rm vis}=\SI{1}{\MeV}$ for the position.
\end{abstract}
\flushbottom


\section{Introduction}
\label{sec:intro}
\subfile{01_intro.tex}

\section{Dataset}
\label{sec:dataset}
\subfile{02_dataset.tex}

\section{Architectures of Machine Learning Models}
\label{sec:architectures}

\subfile{03_BDT}

\subfile{03_DNN}

\subfile{03_CNN}

\subfile{03_GNN}

\section{Results}
\label{sec:results}

\subfile{04_definitions}

\subfile{04_vertex}

\subfile{04_energy}

\subfile{04_comp_time}

\section{Discussion}
\label{sec:discussion}

\subfile{05_discussion}

\section{Conclusions}
\label{sec:conclusion}
\subfile{06_conclusions}

\section{Acknowledgements}

\subfile{07_acknowledgement}


\printbibheading[heading=bibintoc]
\printbibliography[resetnumbers=true,heading=none]

\clearpage

\end{document}

%% file: 01_intro.tex
JUNO is a \SI{20}{\kilo\tonne} Liquid Scintillator (LS) detector designed with the primary goal of
determining the neutrino mass ordering and the precise measurement of the neutrino oscillation
parameters $\sin^2 \theta_{12}$, $\Delta m^2_{21}$ and $\Delta m^2_{31}$~\cite{An:2015jdp}.
JUNO is situated at an equal distance of about \SI{52.5}{\kilo\m} from two nuclear power plants
and registers neutrinos produced in reactor cores via the neutrino inverse beta-decay (IBD) channel: \inversebetadecay.

The detailed description of the JUNO detector and the physics programme may be found
in~\cite{Adam:2015vqa,An:2015jdp}. The detector design is typical for the experiments with
reactor electron antineutrinos. Liquid organic scintillator, rich with protons, is both the target for
the IBD reaction and the detection medium.
When excited by the interaction with a charged particle, the scintillator produces visible light
with a yield roughly proportional to the particle energy. It is also
able to indirectly detect neutral particles. For instance, neutrons are captured by hydrogen atoms, producing \SI{2.2}{\MeV} de-excitation gammas, which can ionize other atoms, liberating electrons that are visible to the scintillator.
Thus, the two-fold signal of the positron and the neutron from IBD provides a clear signature for a neutrino interaction. In this process the most of the neutrino energy is transferred to the positron.

\begin{figure}[htpb]
  \centering
  \adjincludegraphics[Clip=0 0 0 {0.404\height},width=0.7\linewidth]{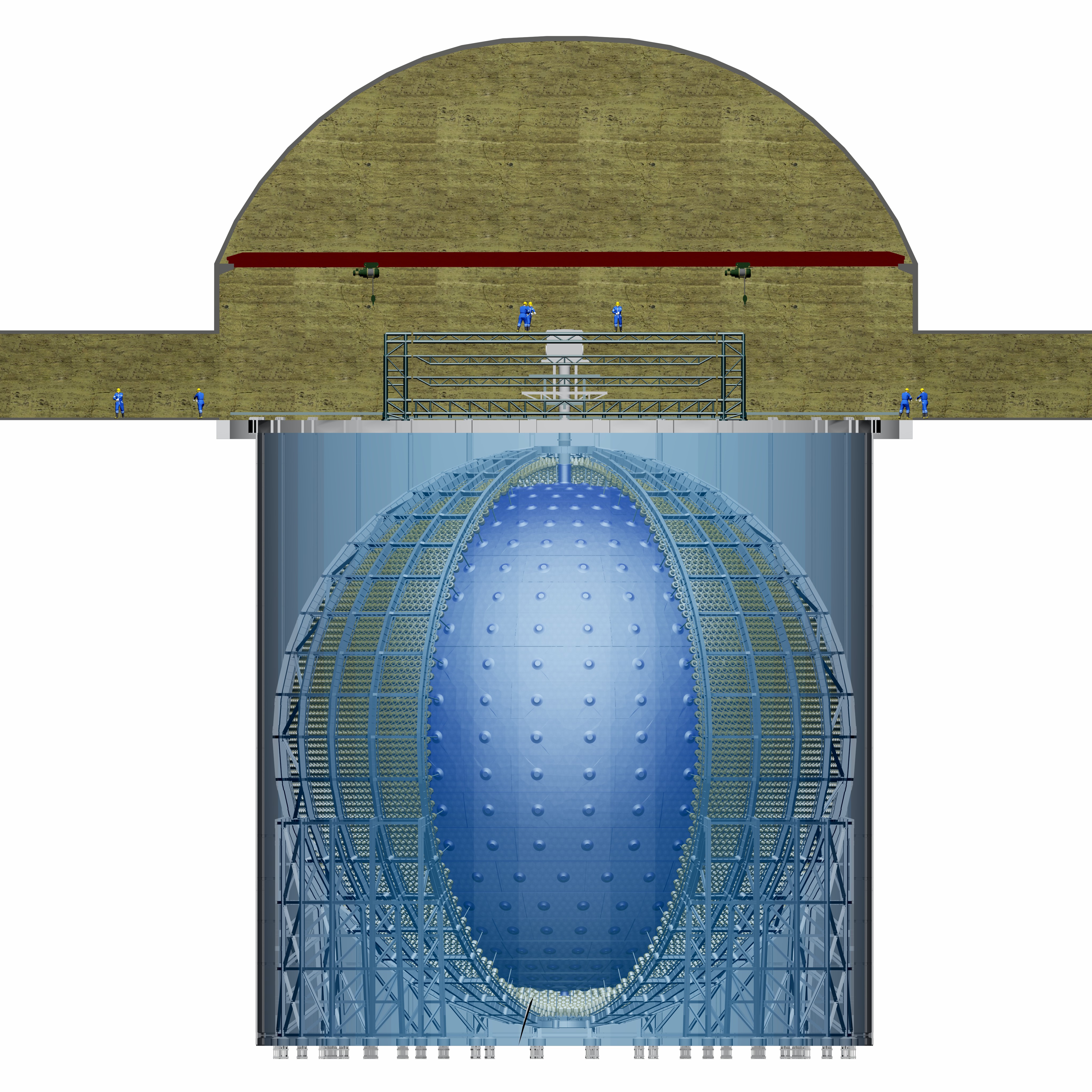}
  \caption{The central detector of JUNO. An acrylic sphere filled with 20 kt of liquid scintillator is
  located in a water pool. Around \num{17600} large PMTs looking inside are installed on a steel
  supporting structure.}%
  \label{fig:juno-cd}
\end{figure}

The detector design is presented in Figure~\ref{fig:juno-cd}. The target, consisting of \SI{20}{\kilo\tonne} of LS, is contained in a
transparent acrylic sphere with an inner diameter of \SI{35.4}{m}, which is located in a cylindrical water pool.
Photosensitive detectors facing inwards --- almost \num{17600} large
$20''$ photo-multiplier tubes (PMT) --- are installed on a steel structure around the acrylic sphere.
The geometrical coverage of 75\% is close to the maximum. The detector is also equipped with
\num{25600} small $3''$ photomultipliers (sPMT), placed between the large PMTs, providing an extra coverage of 3\%.
The signal collected by PMTs is used for estimation of the vertex and the energy of neutrino
and background particle interactions.

The goal of the neutrino mass ordering determination at the level of 3 standard deviations 
poses stringent requirements
on the detector performance: the energy resolution width must be better or equal to
$\sigma=3\%$ at $\SI{1}{MeV}$ and the energy nonlinearity uncertainty should be $<1\%$.
The methods traditionally used for the energy reconstruction rely on the reconstructed vertex
and require a spatial resolution of at least 10~cm. Reconstruction of the vertex is also needed
for time-of-flight correction for pulse shape discrimination of particle types.
The list of tasks requiring vertex reconstruction is not limited by the two aforementioned cases.
Previous studies~\cite{An:2015jdp} indicate that the achievable spatial resolution is of
the order of \SI{10}{\cm}, with the ideal precision limit being of the order of a few centimetres.


On a technical level, a high energy resolution is achieved in JUNO by maximizing the light collection
via optimization of the LS composition, the usage of PMTs with high quantum efficiency, and large
geometrical coverage. The energy nonlinearity uncertainty will be controlled by a calibration
campaign~\cite{JUNO_calib}.
On a software level the precision of the energy reconstruction procedure should match the
requirements as well.


In recent decades Machine Learning (ML) is becoming more and more popular in high energy
physics~\cite{Baldi:2014kfa} both in collider experiments~\cite{Guest:2018yhq} and in neutrino
experiments~\cite{terwilliger2017vertex}. In addition to signal and background
separation~\cite{Shirobokov:2020tht}, we can also find its broad application in jet and
event reconstruction~\cite{Guest:2016iqz}.
ML has a long history of development. The concept of the perceptron, the basic unit of current shallow networks, was proposed in 1958, and it is now known as a Rosenblatt's
perceptron~\cite{rosenblatt1958perceptron}. Backpropagation, an algorithm that is used to train
almost any neural network, was proposed in the 1980s~\cite{rumelhart1986learning}. However, after
that, the interest to deep learning revived only in 2005$-$2009~\cite{bengio2006greedy,raina2009large}
with the rapid development of computational technologies, like GPUs, and collection of the
significant amounts of data needed to train deep networks.

Given the tremendous success of ML in high energy physics, we investigate ML-based approaches
for energy and vertex information in the JUNO experiment. The time and charge data collected
from the PMTs can be provided as input for supervised training of ML models.
We observe that, after training on many samples, networks can predict the event vertex and energy with high quality and fast speed.

In this paper, we study the applicability of different ML approaches to the vertex
and the energy reconstruction for the energy range of \SIrange{1}{10}{\MeV} covering
the region of interest for IBD events from reactor electron antineutrinos.
The dataset used for the training of the models is introduced in Section~\ref{sec:dataset}.
In Section~\ref{sec:architectures} we briefly summarize the main concepts and terminology
of utilized ML approaches and describe the details of the models adopted for our task.
The performance of these models is presented in Section~\ref{sec:results}, where we also discuss
the impact of the electronics effects on the reconstruction accuracy. The applicability of ML
to real data is discussed in Section~\ref{sec:discussion}. This section also covers a few future
directions of work.
Finally, conclusions are given in Section~\ref{sec:conclusion}.

%% file: 02_dataset.tex
\subsection{Data preparation}

The training and testing of neural networks has been performed on Monte Carlo 
samples generated with the official JUNO software~\cite{JUNO_software} and further 
processed to include the most relevant effects of the electronics response.

The detector simulation is based on the Geant4 framework~\cite{Geant4} with
the geometry~\cite{Li:2018fny,zhang2020method} implemented in details according to the latest design. 
The simulation starts from the injection of positrons of different energies 
in the range of \num{0}-\SI{10}{\MeV} characteristic for the IBD reaction
induced by reactor neutrinos.

After the generation of the primary particles, their ionisation energy losses are simulated. 
In the liquid scintillator, this process is accompanied by the production of scintillation and 
Cerenkov photons. The majority of photons in the region of PMT sensitivity are 
produced by scintillation, which yields $O(10^4)$ photons per \si{\MeV} of deposited 
energy. 
The simulation also includes interactions of primary particles with other particles 
of the medium, and allows tracking the secondary particles as well. 
In particular, after depositing its kinetic energy $E_{\rm kin}$, the positron 
annihilates with an electron, 
producing a pair of gammas with energies of \SI{511}{\kilo\eV} each. 
These annihilation gammas then ionize the medium, causing the production of the scintillation light.
If gammas appear close to the detector edge, they may 
escape from the detector, carrying away part of the energy, which leads to lower light yields 
and complicates the reconstruction. 

The photons are propagated through the detector with the most relevant optical processes 
taken into account. The photons reaching the PMT photo-cathode may then 
produce a photo-electron (p.e.) according to the photon detection efficiency 
measured in the laboratory~\cite{JUNO_PMT}. About \num{1350} p.e. are detected per 
\SI{1}{\MeV} of deposited energy at the detector center. This number is affected by statistical fluctuations and 
systematic effects like LS non-linearity and detector response non-uniformity. 
The event of p.e. detection is called a hit. 
The charge taken from PMT is, at the first order, proportional to the 
number of hits. In this work we ignore any statistical and systematic deviations from 
the linear relationship between the number of hits and the charge, and refer to the 
number of hit on PMTs as a charge information. It is hard to measure the time of
each hit, instead we only assume that the first hit time on each PMT is measured.
The time information is counted from the time of event generated in simulation.

The second part of the data preparation, the application of electronics response effects, 
is done on top of the data produced by the detector simulation. Two effects
are included: the dark noise (DN), i.e.\ spontaneous hits appearing in PMTs; 
and the transit time spread (TTS), which happens due to stochasticity of the photo-electron 
path from the photo-cathode to the anode. 
In this study, only $20''$ PMTs from the central detector are used for the reconstruction, 
including around \num{5000} Hamamatsu dynode PMTs (R12860) and around \num{12600} Micro Channel Plates (MCP) 
PMTs from North Night Vision Technology (NNVT). 
The information on the TTS distributions was taken from factory parameters of PMTs,
\SI{2.6}{ns} for Hamamatsu and \SI{19}{ns} for NNVT,
with additional smearing applied according to the measurements performed at Daya Bay
neutrino experiment (predecessor of JUNO). 
The DN rates are sampled from distributions based on the test measurements performed 
for a sample of PMTs expected to be installed in the central detector of JUNO. The 
distributions span up to \SI{50}{kHz} for Hamamatsu and up to \SI{100}{kHz} for NNVT.


In this work, reconstruction is based exclusively on the data from the $20''$ PMTs, 
since the area coverage of the $3''$ PMTs is significantly smaller.

\subsection{Data structure}

We have prepared the training and the testing datasets with the following settings and statistics:

\begin{enumerate}
    \item {\bf Training dataset} consists of 5 million events, uniformly distributed in kinetic energy from $0$ to \SI{10}{\MeV} and in the volume of the central detector (in liquid scintillator). Typically, only the first 90\% of this dataset is reserved for iteratively optimizing the models' parameters. In-between each training pass over this data (epoch), the models' performance is validated over the last $10\%$ of the dataset.
    \item {\bf Testing dataset} consists of subsets with discrete kinetic energies of \SI{0}{\MeV}, \SI{0.1}{\MeV}, \SI{0.3}{\MeV}, \SI{0.6}{\MeV}, \SI{1}{\MeV}, \SI{2}{\MeV}, ..., \SI{10}{\MeV}. Each subset contains 10 thousand events. This dataset is used to estimate performance after the end of training.
\end{enumerate}

The targets of the reconstruction task are the following two variables: the total energy 
deposited by each positron and by-product gammas inside the LS ($E_{\rm dep}$) and the 
average position of the energy deposition calculated as:
\begin{align}
  \mathbf{r}_{\rm dep} = \frac{1}{N_{\rm s}}\sum_i^{N_{\rm s}} \mathbf{r}_{{\rm s},i} E_{{\rm dep}, i}, 
\end{align}
where $\mathbf{r}_{{\rm s},i}$ is the position of the $i$-th simulation step and $E_{{\rm dep},i}$ 
is the energy deposited at this step. The summation runs over all the steps.
These two variables ($E_{\rm dep}$ and $\mathbf{r}_{\rm dep}$) are also referred to as 
true information and are available 
for training from the Monte Carlo simulation. They are predicted at the testing stage and 
the predictions are compared to the true values.

As inputs, we use two types of information: aggregated features precalculated from 
PMT signals and PMT-wise information. 

The former is used for the simpler models, and are intuitively chosen to be the ones which could contain
the most information needed for reconstruction:

\begin{enumerate}
  \item The total number of detected photo-electrons (hits), which is at first order proportional 
to the deposited energy. This number is less than the number of photons reaching the photo-cathode of PMTs
because of the limited $\sim$30\% detection efficiency.
\item The center of charge, defined as
\begin{align}
  \mathbf{r}_{\rm cc} = \frac{1}{N_{\rm hits}}\sum_i^{N_{\rm PMTs}} \mathbf{r}_{{\rm PMT}_i} n_{\rm p.e.},
\end{align}
which is a rough estimation of the energy deposition location and allows accounting for the 
non-uniformity of the detector response. The summation runs over all fired PMTs with 
positions $\mathbf{r}_{{\rm PMT}_i}$ and numbers of detected photo-electrons $n_{\rm p.e.}$.
  \item Mean and dispersion of the first hit time distributions. These two features provide extra information 
about the dynamics of the signal.
\end{enumerate}

\begin{figure}[!htb]
    \centering
    \includegraphics[width=0.39\textwidth]{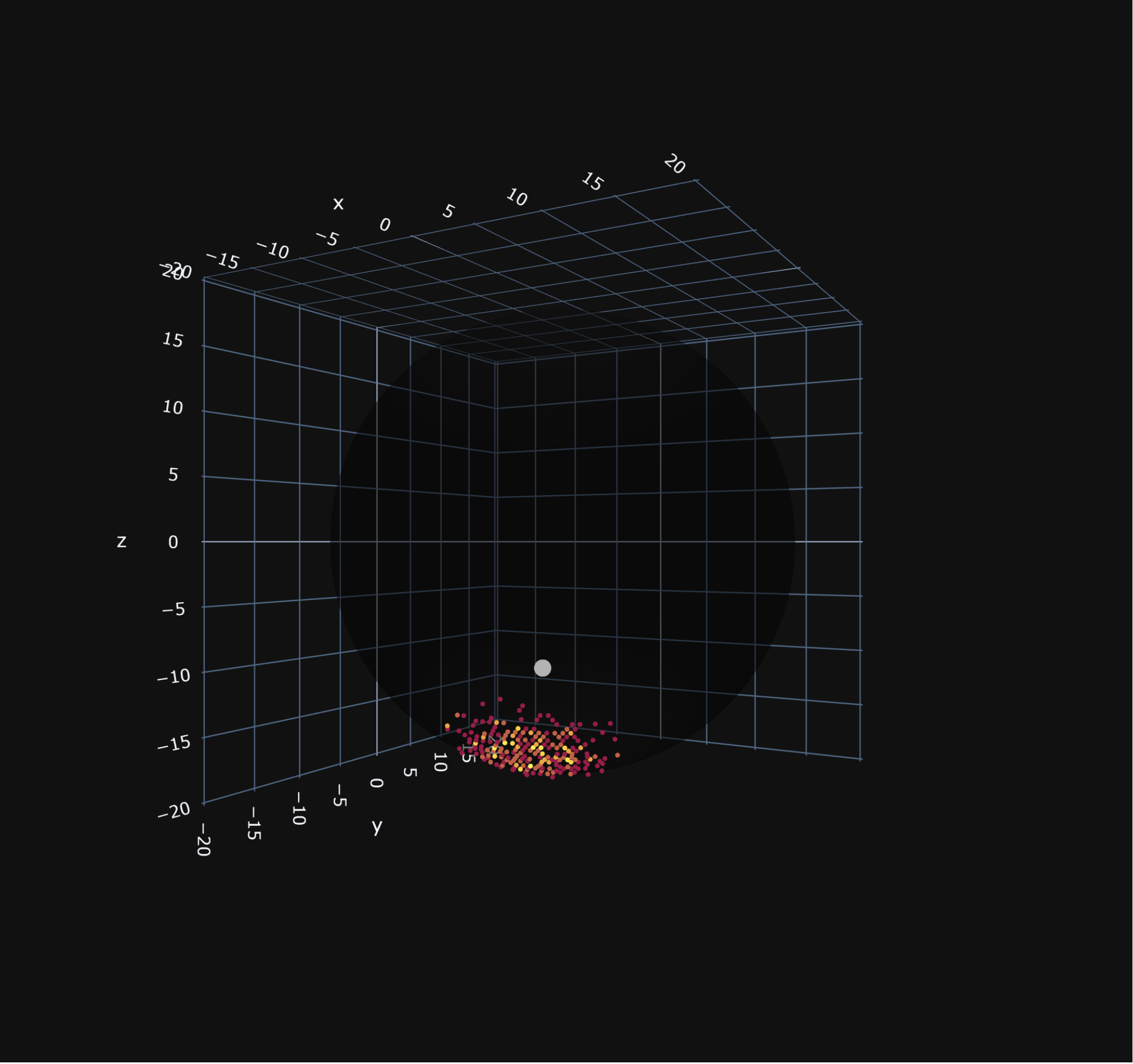}
    \includegraphics[width=0.39\textwidth]{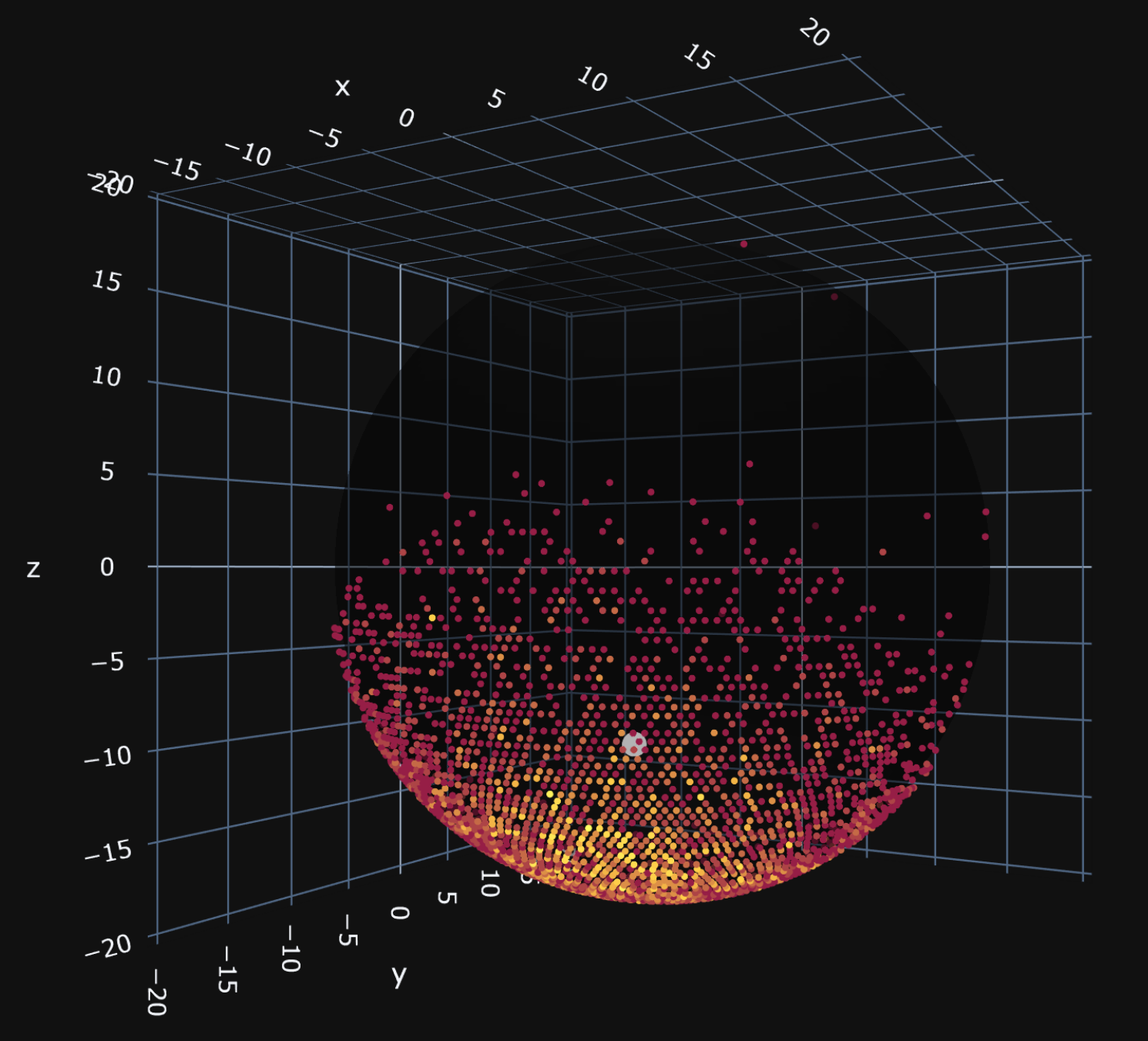}\\
    \includegraphics[width=0.39\textwidth]{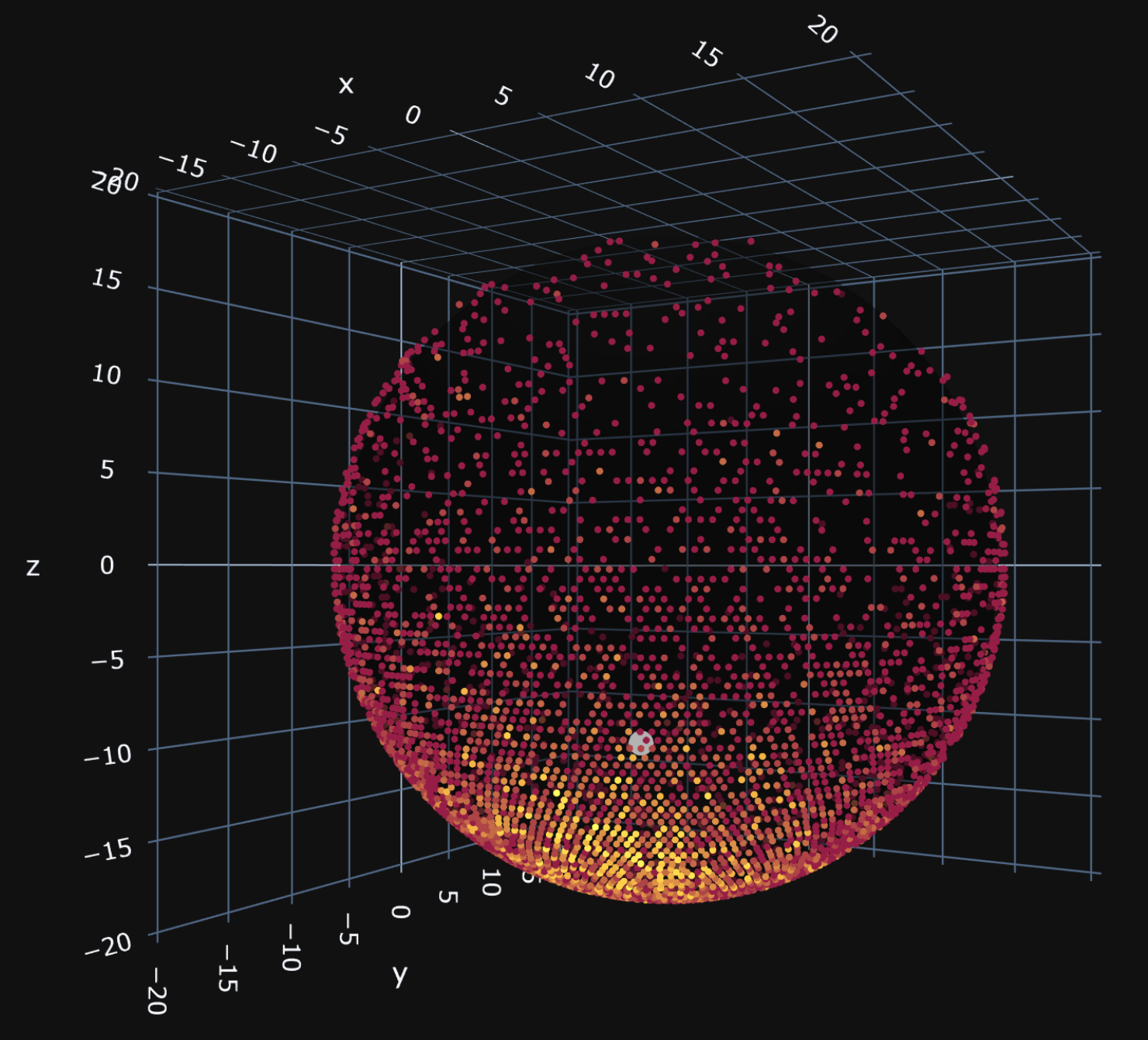}
    \includegraphics[width=0.39\textwidth]{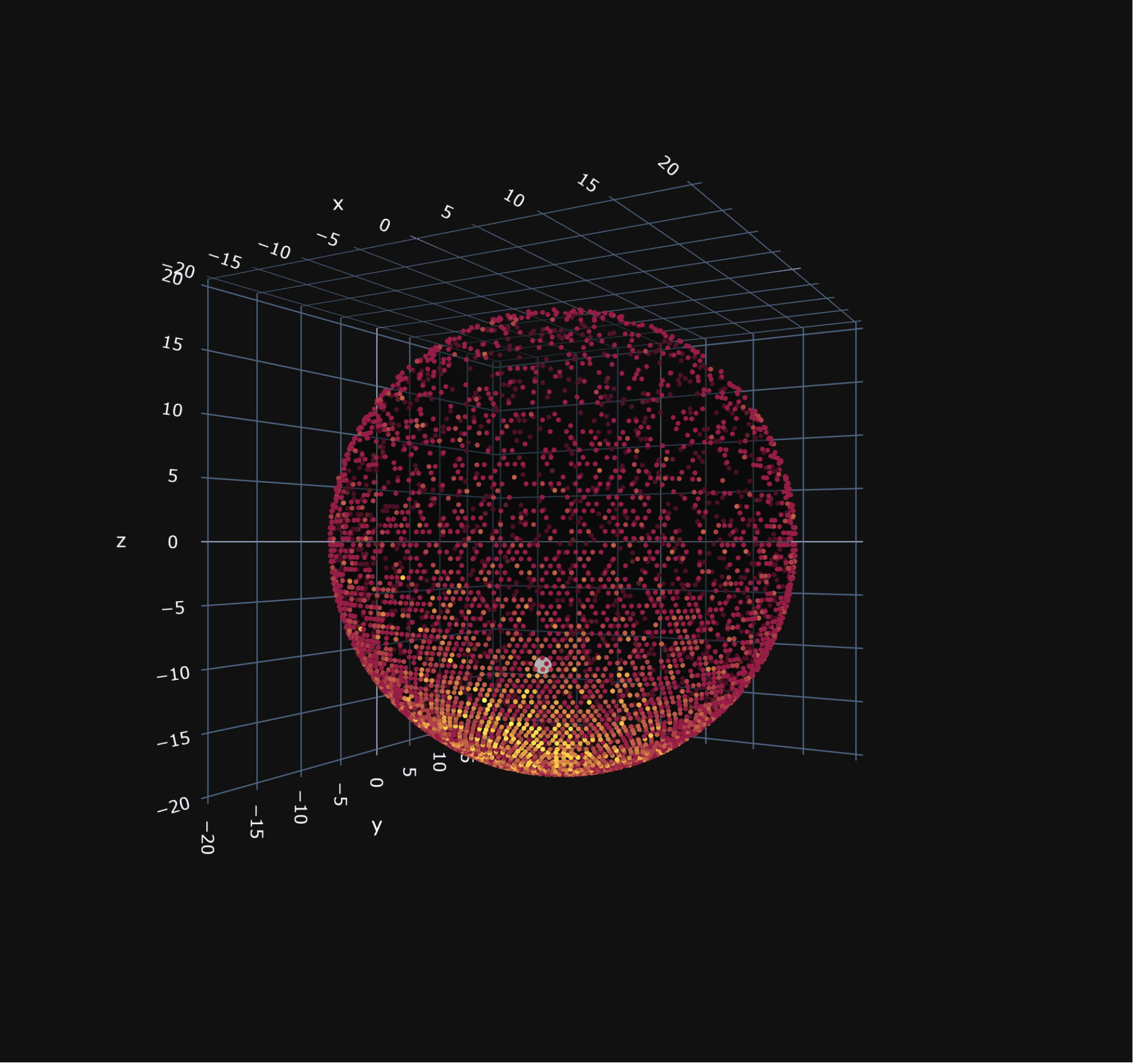}\\
    \caption{Example of an event seen by PMTs and evolving in time: 
    5th~ns (top left), 55th~ns (top right), 105th~ns (bottom left) and 155th~ns (bottom right).
    Only fired PMTs are shown. The color represents the accumulated charge in PMTs:
    yellow points show the channels with more hits, red points --- the channels with few hits.
    Only hits in $20''$ PMTs are shown. The primary 
    vertex is shown by the gray sphere.}
    \label{fig:data_vis}
\end{figure}

The PMT-wise measured information is used as input for the more complex models, and includes the 
number of hits and the time of the first hit at each PMT. Figure~\ref{fig:data_vis} illustrates the time evolution of the signal in the PMT channels for 
a positron event of \SI{5.5}{\MeV}. 
The event display software~\cite{You:2017zfr,Zhu_2019} dedicated to JUNO can be used to dynamically display the entire process.
All the inputs are summarized in Table~\ref{tab:data_structure}. 

\begin{table}[!htb]
    \centering
    \begin{tabular}{lrr}
        \toprule
        Parameter                                                  & Name                                                  & Type [$\times$size] \\
        \midrule
        \multicolumn{2}{c}{\it True information}                  \\

        Event ID                                                   & int                                                  \\
        Deposited energy                                           & \texttt{Edep}                                         & float                \\
        Average position of the energy deposition                  & \texttt{x\_edep}, \texttt{y\_edep}, \texttt{z\_edep}  & float $\times$ 3     \\
        \midrule
            \multicolumn{2}{c}{\it Aggregated information}        \\
        Total number of hits                                       & \texttt{nHits}                                        & int                  \\
        Center of charge coordinates                               & \texttt{x\_cc}, \texttt{y\_cc}, \texttt{z\_cc}
                                                                   & float $\times$ 3                                     \\
        Radial component of center of charge                       & \texttt{R\_cc}                                        & float                \\
        Average of the first hit time                              & \texttt{ht\_mean}                                     & float                \\
        Dispersion of the first hit time                           & \texttt{ht\_std}                                      & float                \\
        \midrule
            \multicolumn{2}{c}{\it PMT-wise measured information} \\
        Number of  hits (photoelectrons)                           & \texttt{npe}                                          & int                  \\
        Hit time of the first detected photon                      & \texttt{hittime}                                      & float                \\
        Position                                                   &                                                       & float $\times$ 3     \\
        Type                                                       & \multicolumn{2}{r}{$20''$ Hamamatsu / $20''$ NNVT}       \\
        \bottomrule
    \end{tabular}
    \caption{Data structure of a single event. See details in the text.}
    \label{tab:data_structure}
\end{table}

To study the impact of TTS and DN on the reconstruction, we have generated 
four datasets for training and four corresponding testing datasets with different 
TTS/DN options. The first dataset does not include the influence of TTS and DN. 
The other two datasets only include the influence of either TTS or DN. Then, the last dataset 
includes both TTS and DN, which is the situation closest to the real case, and is
used as the default dataset in the following study if not specified otherwise.

%% file: 03_BDT.tex
The primary vertex and the energy are reconstructed by two classes of models. 
Simple models, that use only several aggregated features, include Deep Neural Networks (DNN) and Boosted Decision Trees
(BDT). The complex models, that use the PMT-wise information include Convolutional Neural Networks (CNN) and Graph
Neural Networks (GNN).

DNN and BDT are minimalistic, which allows getting rough predictions at a very low computational cost.
The CNNs (VGG, ResNet) and GNN are more complex and provide better precision by processing the full information,
at the cost of an increased number of intrinsic parameters and, as a consequence, a slower prediction rate. 

In the current section we discuss the general terminology of supervised learning and the architecture of each model.

\subsection{Boosted Decision Trees}

A decision tree~\cite{dtrees} is a model that consists of a binary set of splitting rules based on values of different features of the object. As a result, the entire feature phase space is split into separate cells. The predicted value for a given object is then completely defined by the cell which the combination of features of the object belongs to.

The decision tree thus may be used as a non-parametric model to be trained by the supervised Machine Learning algorithm. This model is simple and interpretable.
It is common, instead of using a single decision tree as a complete model, to use an aggregated model like an ensemble of decision trees.
Two most common approaches for decision tree ensembles are Boosted Decision Trees~\cite{gb1, gb2} and Random Forests~\cite{rf}.

\begin{figure}[!htb]
	\centering
	\includegraphics[width=\textwidth]{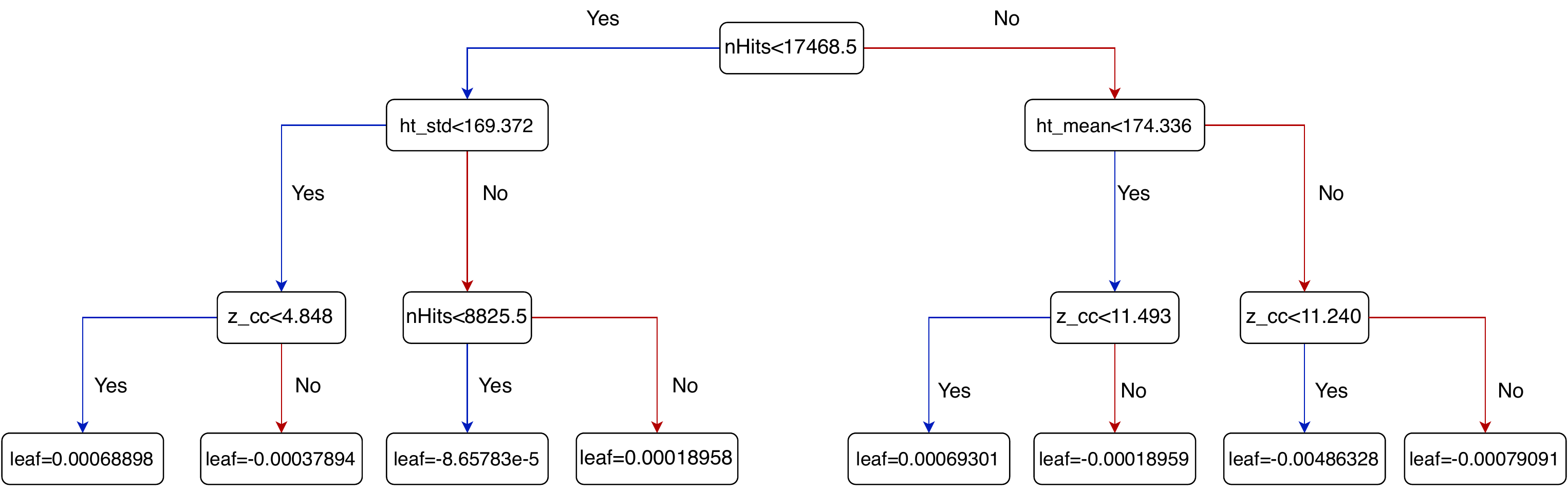}
	\caption{Example of a decision tree with a maximum depth equal to four.}
	\label{fig:DT}
\end{figure}

Random Forests and Boosted Decision Trees differ primarily in how an ensemble model is built. In the former method, $N$
decision trees are trained independently. The final solution for the entire ensemble is a mean prediction. In the latter
method the models are trained sequentially. Each subsequent decision tree is trained to correct errors of all previous
decision trees in the ensemble. Figure~\ref{fig:DT} shows an example of a decision tree from an ensemble of Boosted Decision Trees. The value on the leaf is added to the prediction value of the previous ensemble as a correction.



We explore the same aggregated features listed in Table~\ref{tab:data_structure}.
This allows for a direct comparison of  performances for both approaches.
The XGBRegressor BDT implementation from the XGBoost library~\cite{xgb} is used for this study.


%

\subsubsection{Tuning Boosted Decision Trees for JUNO}

The selection of the optimal hyperparameters was carried out using a grid search. It defines a grid
of the parameter values and calculates the value of an objective function for each node in the grid. 
We use the Root Mean Squared Error (RMSE) as an objective function. The node with the lowest RMSE value
is taken as the optimum.

The grid search was performed for the BDT method with a learning rate of
0.08 trained on the dataset with 1M events, which includes the effects of TTS and DN. We used
cross-validation with the dataset split in three parts with the KFold tool from the Scikit-learn
library~\cite{scikit}. The training of the model was stopped in case the RMSE on the validation set
did not decrease for five consecutive iterations. The resulting RMSE metric was calculated on the
additional test sample with 150K events. 

\begin{figure}[!htb]
	\centering
	\includegraphics[width=0.7\textwidth]{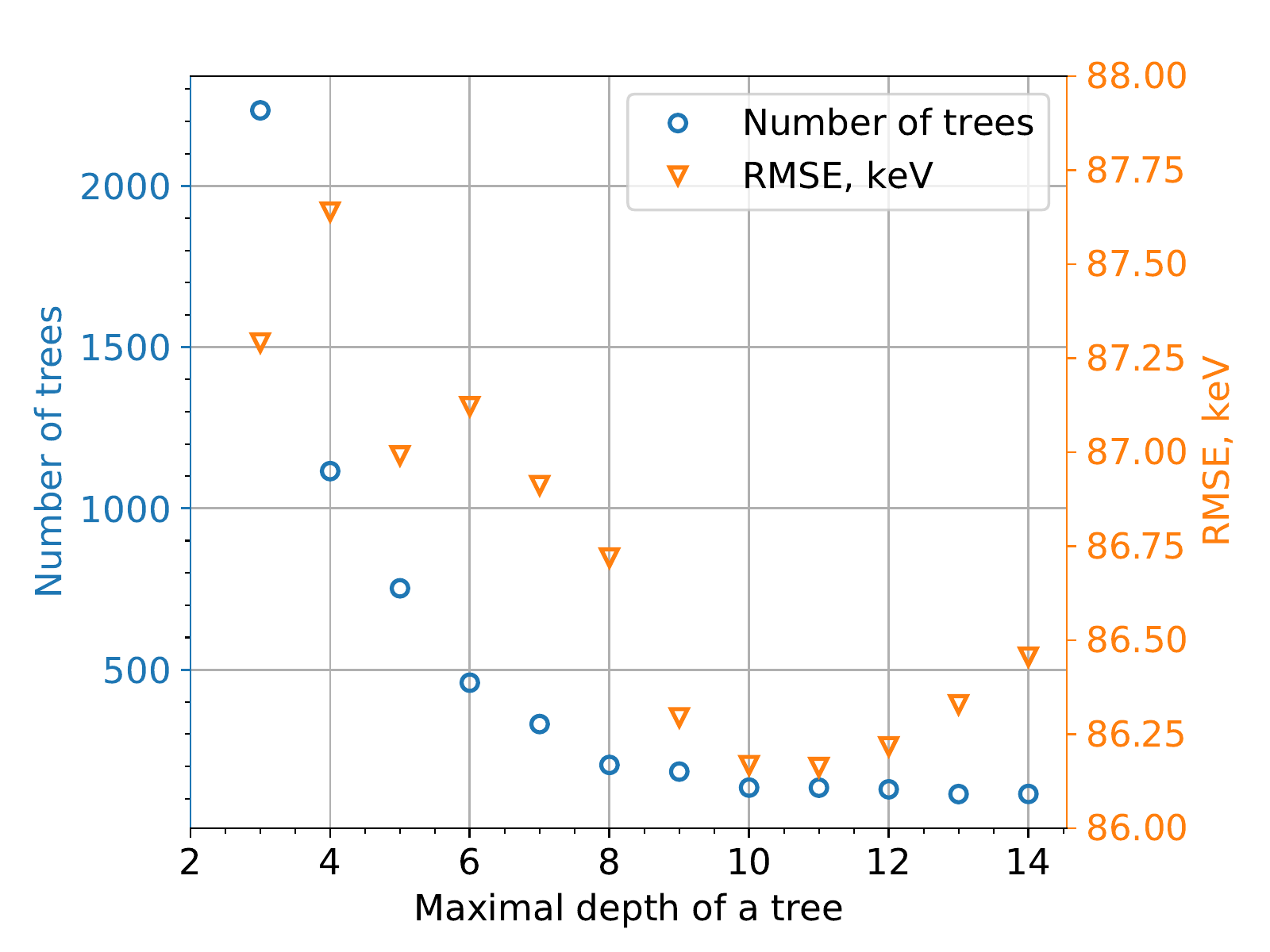}	
	\caption{The dependence of number of trees in the ensemble and RMSE on the test sample on the maximum depth.}
	\label{fig:grid_search_BDT}			
\end{figure} 

The results, presented in Figure~\ref{fig:grid_search_BDT}, demonstrate that the optimal depth of
the trees is 10 or 11. Shallower trees necessitate more trees and yield worse precision. On the 
other hand, deeper trees require fewer trees but provide worse precision because of overfitting.
 
Table~\ref{tab:bdt_perm} shows the permutation importance for the prediction of 
the $z$-coordinate of the vertex and the energy of the event. Permutation importance 
indicates how the prediction error increases when a feature is not available. These weights are computed  
with ELI5 library~\cite{eli5}. Not surprisingly, the most
informative feature for a vertex coordinate is the corresponding coordinate of 
the center of charge, and the total number of hits is the most informative for 
the energy reconstruction. 
Table~\ref{tab:bdt_perm} also shows the other features sorted by their importance.
The order of the other features is similar for both reconstructed variables. 

\begin{table}[!htb]
    \hspace*{\fill}
	\begin{tabular}{lr}
		\toprule
        Feature                               & Weight              \\
        \midrule
        \texttt{z\_cc}                        & 1.999 $\pm$ 0.008 \\
        \texttt{ht\_mean}				      & (2.436 $\pm$ 0.012)$\times 10^{-2}$ \\
        \texttt{R\_cc}				          & (0.156 $\pm$ 0.002)$\times 10^{-2}$ \\
        \texttt{ht\_std}                      & (5.720 $\pm$ 0.028)$\times 10^{-4}$ \\
        \texttt{nHits}				          & (1.460 $\pm$ 0.004)$\times 10^{-4}$  \\
		\bottomrule
		\multicolumn{2}{c}{\texttt{z\_edep}} 
    \end{tabular}
    \hspace*{\fill}
	\begin{tabular}{lr}
		\toprule
        Feature                            & Weight              \\
	    \midrule 
        \texttt{nHits}                     & 2.041 $\pm$ 0.005 \\
		\texttt{ht\_mean}			       & (1.365 $\pm$ 0.002)$\times 10^{-2}$ \\
		\texttt{R\_cc}	   				   & (1.049 $\pm$ 0.005)$\times 10^{-2}$ \\
		\texttt{z\_cc}      			   & (1.900 $\pm$ 0.006)$\times 10^{-3}$ \\
		\texttt{ht\_std}				   & (1.623 $\pm$ 0.005)$\times 10^{-3}$ \\
		\bottomrule
		\multicolumn{2}{c}{\texttt{edep}} 
	\end{tabular}
    \hspace*{\fill}

	\caption{Permutation importance for the prediction of the $z$-coordinate (left)
     and the deposited energy (right) on 1 million events.
}
	\label{tab:bdt_perm}
\end{table}

%% file: 03_DNN.tex
\subsection{Deep Neural Networks}

A Neural Network is a method to construct the function $f_{\mathrm{W}}$ in a modular fashion, inspired by the structures
of a human brain.
The simplest network consists of a single unit, called a neuron, which computes a linear combination of its inputs and passes it to a non-linear activation function $h$:
\begin{align}
    f_{\mathrm{W}}(\bm{x}) = h(\mathrm{W} \bm{x} + \bm{b}),
\end{align}
where the offset $\bm{b}$ is called the bias of a neuron. There are different methods to define the activation function $h$.
The most popular one is the Rectified Linear Unit (ReLU)~\cite{relu}, defined as $h(y) = \max(0, y)$.

\begin{figure}[H]
    \centering
    \includegraphics[width=.7\textwidth]{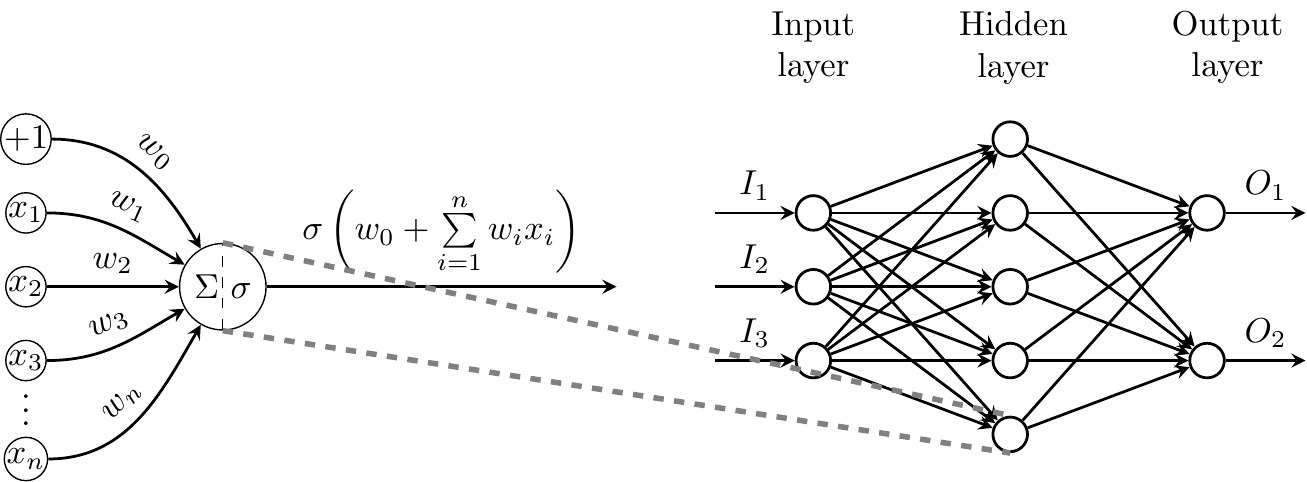}
    \caption{Representation~\cite{MLP-plot} of a simple Neural Network with $3$ layers.}
    \label{fig:MLP}
\end{figure}

Multiple neurons are combined in a layered structure, forming a Neural Network (see Figure~\ref{fig:MLP}).
Layers in which all neurons are connected to the previous units are said to be fully connected, or dense.
When the layers structure go deeper, the network becomes complex and have better ability at fitting, which forms a Deep Neural Network (DNN).

\begin{figure}[H]
    \centering
    \includegraphics[width=.5\textwidth]{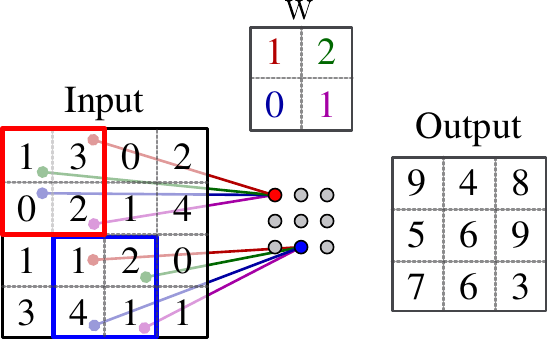}
    \caption{2D convolution of a $4\times 4$ image. The receptive fields of two neurons (red and blue squares) are
    shown. Note that they are of the same shape and size, which enables them to share the respective weights, forming a
   filter W, which is also referred as a kernel. In this way $4 \times 9$ parameters are reduced to $4$.}
   \label{fig:conv}
\end{figure}

When working with high dimensional data, such as images, the connection count grows drastically, and must be reduced to make training more efficient.
To do so, the neurons are linked only to a subset of the previous units, which forms their receptive field. Convolutional
Neural Networks (CNN) take this concept a step further, by making all receptive fields of a layer the same size and
shape, which is usually that of a square patch of input data. 
In this way, the computed linear combination is a discrete convolution (see Figure~\ref{fig:conv}). CNN has a good advantage in information extraction for high-dimensional data. The methods using CNNs will be introduced in the later sections (see Section~\ref{sec:03_CNN}--\ref{sec:03_GNN}).

However, in this section, we would like to discuss the method that implements the basic DNN first.

\subsubsection{Tuning Deep Neural Networks for JUNO}

In theory a DNN can approximate an arbitrary function given it has enough layers and
neurons. 
However, on practice, the efficiency
of the DNNs heavily depends on a choice of the architecture and the optimization
procedure. Thus, an essential part of the application of a DNN to a real-world problem
is an accurate choice of hyperparameters.

In order to understand the achievable precision for such a minimalistic input 
(the aggregated features listed in Table~\ref{tab:data_structure})
and to identify
the optimal architecture we have performed an extensive optimization of the hyperparameters.
The hyperparameters being optimized as well as their respective ranges and optimal choices are 
all presented in Table~\ref{table:dnn_hyperparameters}.

\begin{table}[!htb]
	\centering
	\begin{tabular}{ll}
      \toprule
      Parameter                                                               & Range and \emph{optimum}                          \\
        \midrule
		Activation                                                            & \emph{ReLU}, Tanh                                 \\
        Initialization                                                        & \emph{normal}, orthogonal, uniform                \\
        Scheduler type~\cite{smith2018disciplined,loshchilov2016sgdr}         & ReduceLROnPlateau, \emph{CosineLRAnnealing}, None \\
        Optimizer~\cite{kingma2014adam,zou2019sufficient,mandt2018stochastic} & \emph{Adam}, RMSprop, SGD                         \\
        Layer norm~\cite{ba2016layer}                                         & True, \emph{False}                                \\
        Depth (number of hidden layers)                                       & [2, 10]: \emph{7}                                 \\
        Width  (number of neurons in a layer)                                 & [16, 128]: \emph{32}                              \\
        Batch size                                                            & [128, 1024]: \emph{768}                           \\
        log(lr) (learning rate)                                               & [$-4$, $-2$]: \emph{0.002}                            \\
        \bottomrule
	\end{tabular}
	\caption{Hyperparameters of Deep Neural Networks being tested,
    their ranges and options.
    The values considered as the optimal are marked with \emph{emphasis}.}
	\label{table:dnn_hyperparameters}
\end{table}

We have trained the networks by sampling the hyperparameters randomly. To select the
best hyperparameters, we have performed an empirical posterior estimation of the parameters.
Firstly, we selected 10\% of the top-scoring experiments. Then for the continuous variables we
performed kernel density estimation~\cite{silverman1986density} (KDE) ---
a non-parametric approach for the reconstruction of the data distribution.
According to the procedure of KDE we assigned
a continuous kernel function (in our case, a Gaussian kernel) to each data point in order to
build a smooth approximation
of the density. We selected kernel bandwidth according to the Scott's Rule~\cite{scott2015multivariate}.
For the discrete variables, we counted the probability of each selection.
This procedure gave us a rough estimate of the distribution of hyperparameters of well-performing models.

For the depth and the width, we found two local optima corresponding to a long-and-narrow network and a
short-and-wide network, which is not surprising because the network's complexity depends on both the depth and the width, which are
in some sense interchangeable. However, the interpretation of the depth and the width and
their influence on the network's quality is still an active field of the research. For
example, in~\cite{lu2017expressive} authors argue that depth is responsible for the level of abstraction
that a network operates on, and width is relevant for the flow of the information
through each layer. In our case, based on the results of the posterior estimation procedure, we
have chosen depth equal to 7 and width equal to 32. A similar trade-off is observed for batch size and learning rate,
with the rule of the thumb is that the bigger the batch size implies the bigger learning rate~\cite{NEURIPS2019_dc6a7071}.
We have chosen a batch size equal to 768 and a learning rate equal to 0.002.
As for the other parameters, the choice of their values does not make much difference. Except for a few
combinations of hyperparameters, the vast majority of tested networks provide
comparable performance. The values we have chosen are listed in
Table~\ref{table:dnn_hyperparameters}.

We have also performed an optimization of the loss function for the vertex
reconstruction and for the energy reconstruction by optimizing the coefficients in the
following parametric representation of the combined loss function:

\begin{multline}
\mathcal{L}(g_{\rm true}, g_{\rm pred})  = \alpha \sqrt{\frac{1}{N} \sum_{i=1}^{N} (g_{\rm pred} - g_{\rm true})^2} + \beta \frac{1}{N} \sum_{i=1}^{N} |g_{\rm pred} - g_{\rm true}| + \\
 \gamma \frac{1}{N} \sum_{i=1}^{N}\frac{(g_{\rm pred} - g_{\rm true})^2}{E_{\rm true}} + \delta \frac{1}{N} \sum_{i=1}^{N}\frac{|g_{\rm pred} - g_{\rm true}|}{E_{\rm true}},
\end{multline}
where the weights $\alpha + \beta + \gamma + \delta = 1$ sum to unity, $g$ is a quantity of the interest (energy or coordinates of the vertex).
We note that terms in this loss function have different units, which might affect the optimization of coefficients and
training stability if some loss terms are too big or too small compared to the others.
We found that when energy is represented in GeV units and distance in cm, 
all loss terms have the same order of magnitude.
So, when we reconstruct energy the respective units of coefficients are
$\left\{\alpha\Big[\frac{1}{\mathrm{GeV}^2}\Big], \beta\Big[\frac{1}{\mathrm{GeV}}\Big], \gamma\Big[\frac{1}{\mathrm{GeV}}\Big], \delta\Big[1\Big]\right\}$ and
when we reconstruct vertex the respective units are
$\left\{\alpha\Big[\frac{1}{\mathrm{cm}^2}\Big], \beta\Big[\frac{1}{\mathrm{cm}}\Big], \gamma\Big[\frac{\mathrm{GeV}}{\mathrm{cm}^2}\Big], \delta\Big[\frac{\mathrm{GeV}}{\mathrm{cm}}\Big]\right\}$.

For both the vertex reconstruction and the energy reconstruction we have trained the networks for
different random quadruplets of $\{\alpha, \beta, \gamma, \delta\}$, and we have found that the most optimal
configuration implies nearly equal contribution of these loss components.

\begin{figure}[!htb]
\centering
\begin{subfigure}{0.49\linewidth}
\includegraphics[width=\linewidth]{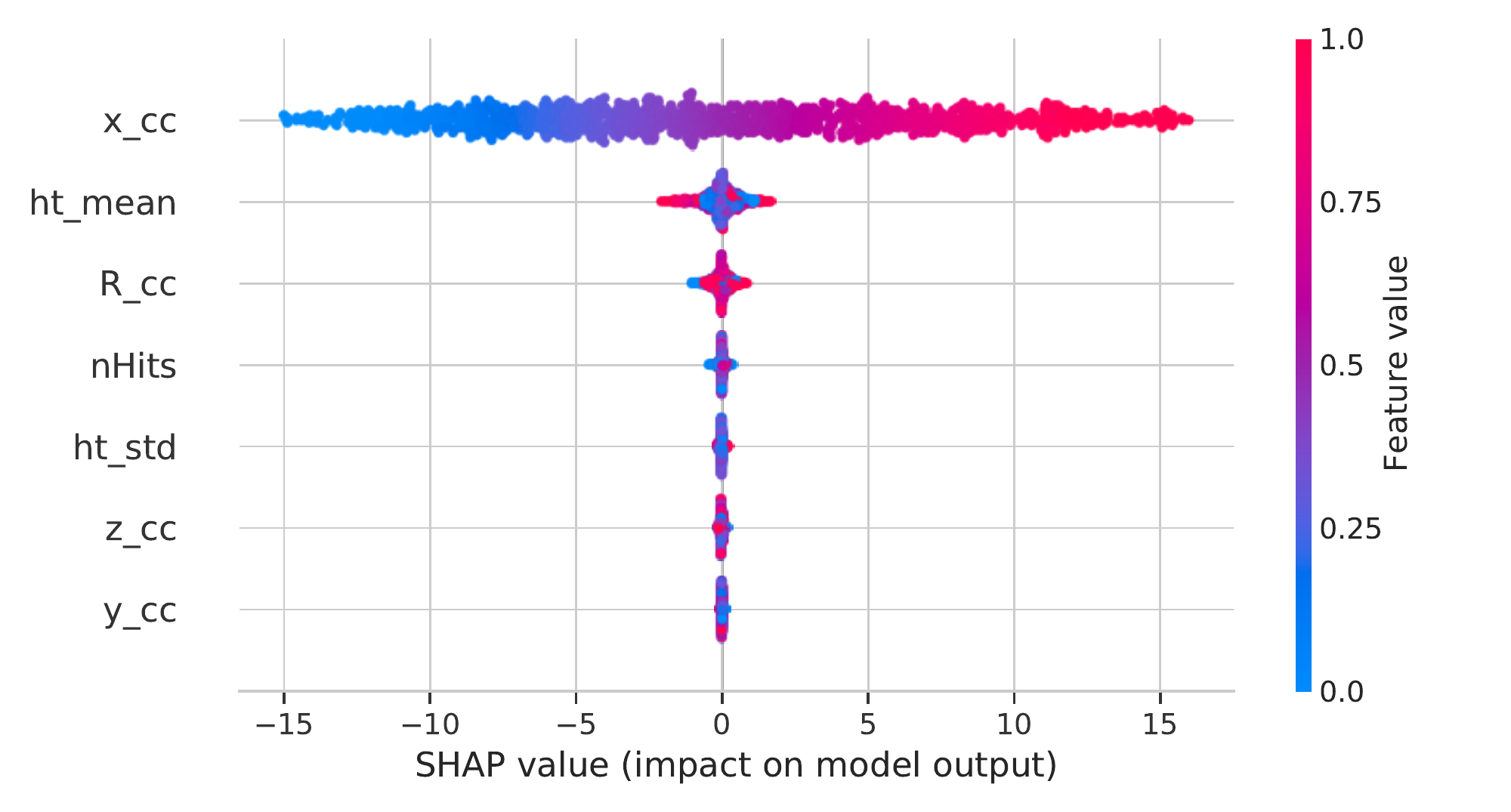}
\end{subfigure}
\begin{subfigure}{0.49\linewidth}
\includegraphics[width=\linewidth]{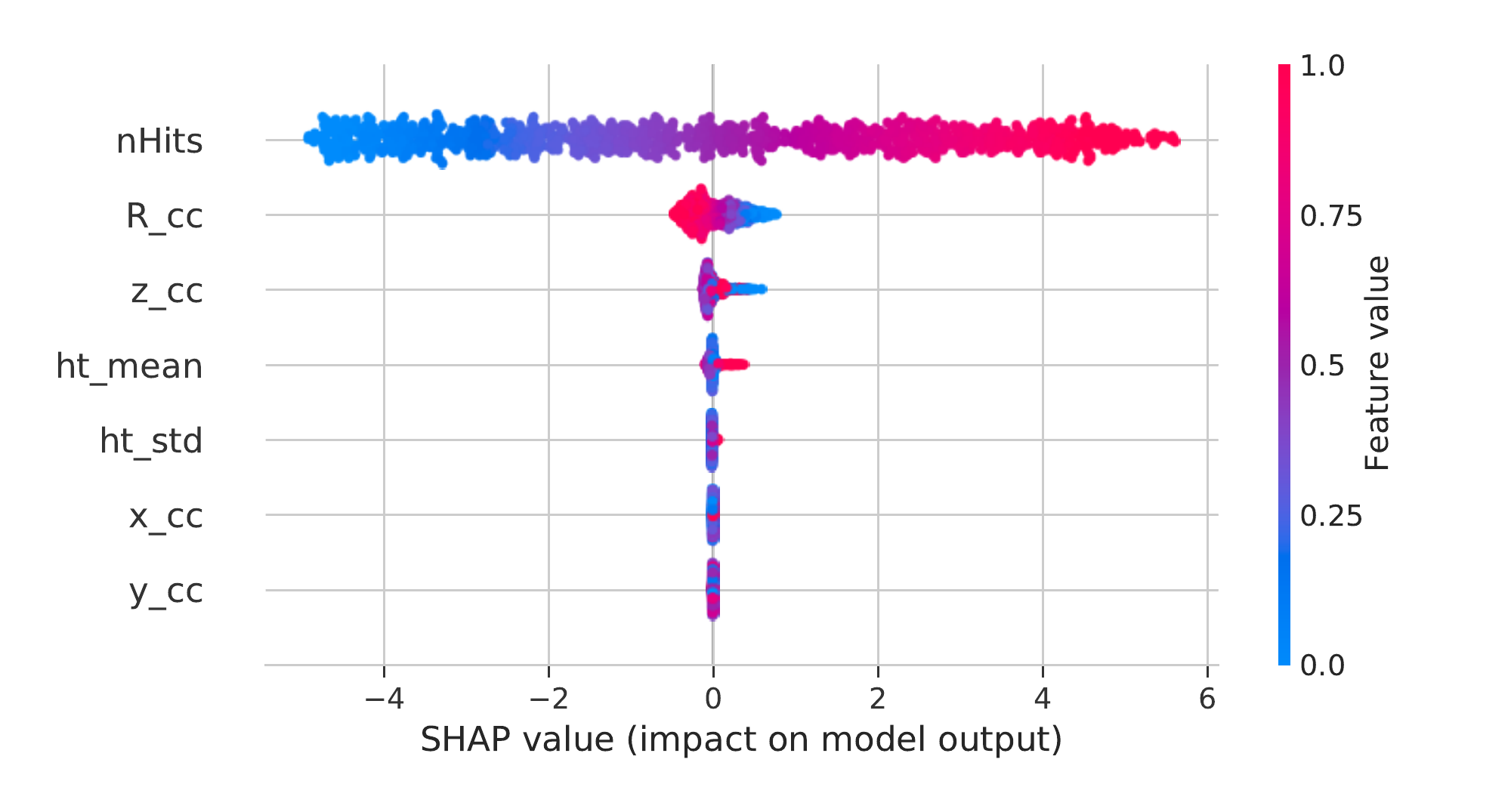}
\end{subfigure}
\caption{SHAP values for the vertex (left) and for the energy (right) prediction.}
\label{fig:shap}
\end{figure}

In order to estimate the importance of each feature, we have computed SHAP (SHapley Additive exPlanations) values~\cite{lundberg2017unified}.
It represents the prediction of the expected model when conditioned on that feature, i.e., shows how
important this single feature with respect to the predicted entity.
In plots, presented on Figure~\ref{fig:shap}, all the features are normalized to the range of [0, 1] and colours represent the normalized features.
For example, we observe, that for the energy prediction,
the most important feature is \texttt{nHits} and an event with a larger number of hits has higher energy predicted by the DNN, which is expected.
The radial component of the center of charge (\texttt{R\_cc}) is the second most important feature
in the energy prediction. It is negatively correlated with the predicted energy.
For the prediction of x-coordinate of the vertex the most important feature is \texttt{x\_cc} and 
the second most important is \texttt{ht\_mean}.

%% file: 03_CNN.tex
\subsection{Planar CNN Models} 
\label{sec:03_CNN}

Convolutional Neural Networks (CNNs) is a deep learning network class commonly applied to analyzing 
visual imagery~\cite{rawat2017deep}.
CNN's architecture makes it feasible to train them on large datasets of images, showing an outstanding 
performance on object recognition and classification tasks. CNNs work 
only on $d$-dimensional Euclidean domains, in which each input sample is represented as a regular grid 
of values, i.e. a $d$-dimensional cubic lattice. In this way, all the neurons of a convolutional layer 
have receptive fields of the same area and shape, capturing the same kind of features, and requiring 
the same number of weights (which is a necessary condition for weight sharing). Geometrically, this 
means that the shape of a receptive field (i.e. of a filter) can be translated to cover different 
\q{patches} of the input data.

The case of using CNNs for reconstruction task in JUNO is to let the network learn the pattern of the 
time and charge distribution obtained from the detector and predict the energy and/or vertex. However, 
PMTs in JUNO are placed along a spherical surface, which is a non-Euclidean domain. No regular grid can 
be constructed on the sphere, meaning that there is no way to define a \q{spherical filter} that can be 
uniformly translated over its surface. Therefore, we need to design a transformation of the spherical 
information obtained from the detector into a planar image.

For CNNs, different network architectures will also affect the final reconstruction performance. We 
tried basic CNN architectures, including AlexNet~\cite{Alex}, VGG~\cite{VGG}, GoogLeNet~\cite{GoogleNet}, 
and so on, and currently get the best performance with VGG and ResNet networks~\cite{Resnet}.

\subsubsection{Projection Method}
\label{ssec:projection}

We have designed a method of projecting on a plane for spherical detectors like JUNO to convert the detector response into a pair of images, one for charge (i.e. the number of PE hits) and one for first hit time. In order to transmit the information more efficiently, the projection design follows the following guidelines:
\begin{enumerate}
    \item A single PMT occupies a single pixel in the image to minimize the information loss.
    \item The final image has a similar arrangement as the original PMT in the detector.
    \item The total amount of pixels should be minimized as soon as the requirement \textbf{1.} is kept. (This is required to reduce the amount of computations.)
\end{enumerate}
There are about \num{17600} PMTs in the central detector of JUNO. If we take a ring of PMTs with same latitude, the higher latitude we take, the fewer PMTs it will contain. To avoid overlap of PMTs, we put the PMTs layer by layer from the top of the detector into the image's pixels so that PMTs with the same latitude will be arranged in the same row. The final image contains $124$ rows, where one row corresponds to the latitude of the PMT. Since the number of PMTs changes with the latitude, the effective number of pixels in each row also changes dynamically. The number of effective pixels per row $N_{\text{eff}}$ and the horizontal position $i_{x}$ of the PMT in the image are given by:
\begin{align}
    N_{\text{eff}} =  \left[N_{\text{max}} \cdot  \frac{\sqrt{R^2-z^2}}{R}\right],  \qquad i_{x} = \left[N_{\text{eff}} \cdot \frac{\arctan(x/y)}{\pi} \right]+ \frac{ N_{\text{max}}}{2},
\end{align}
where $x, y, z$ is the global position of the PMT, $R$ is the radius of the detector and $N_{\text{max}}$ is the total number of columns. $N_{\text{max}}$ controls the tightness of the PMT arrangement. The optimal image size, that ensures that no PMTs are overlapped, is $230\times124$. The image produced by this projection method is similar to that produced by sinusoidal projection, but it is more convenient to manage the position of PMTs in the image. Furthermore, it yields better results compared with the results of the Mercator projection in our study.

\begin{figure}[H]
  \centering
  \begin{subfigure}[b]{0.31\textwidth}
      \centering
      \includegraphics[width=\textwidth]{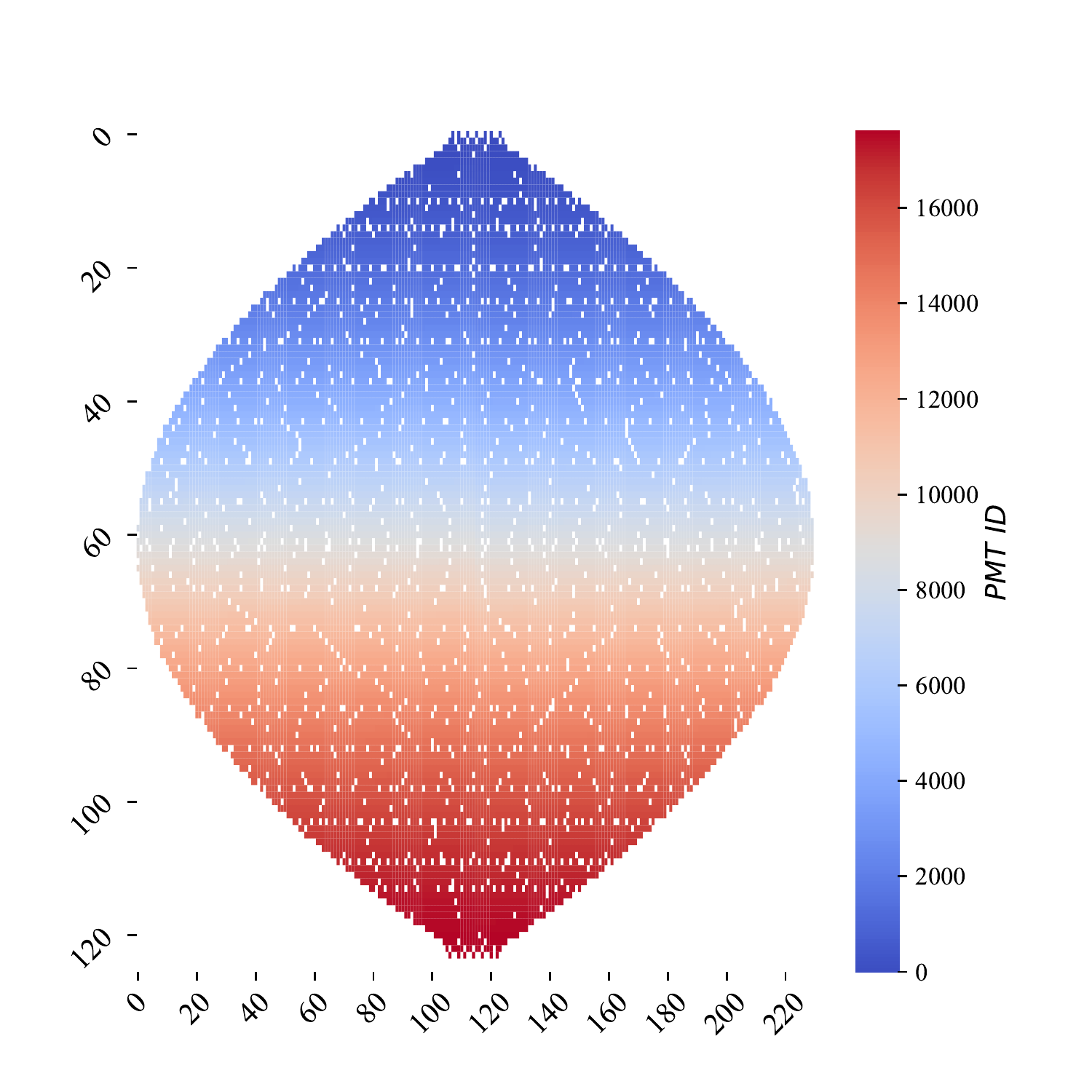}
      \caption{Map of PMTs.}
  \end{subfigure}
  \begin{subfigure}[b]{0.31\textwidth}
      \centering
      \includegraphics[width=\textwidth]{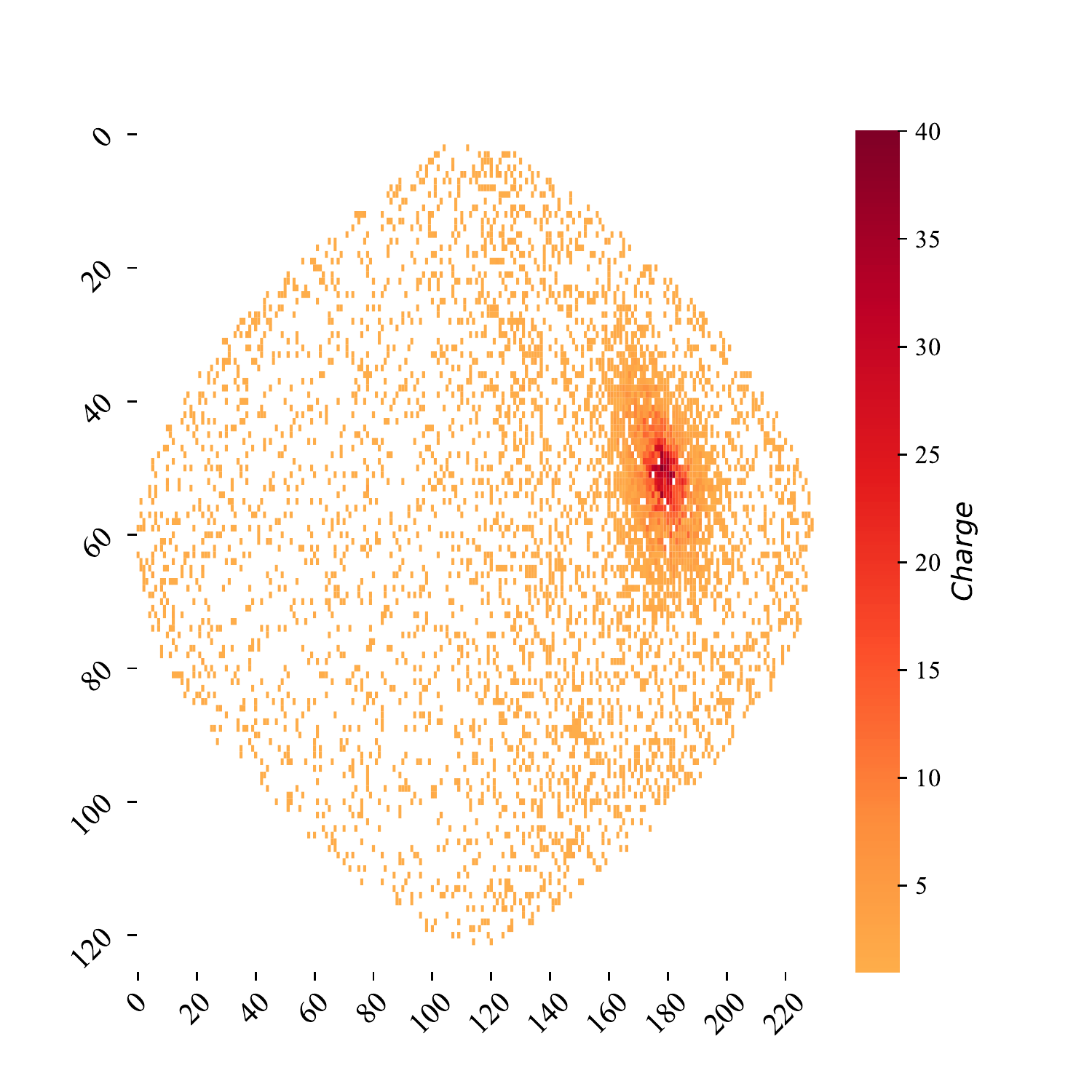}
      \caption{Charge channel.}
  \end{subfigure}
  \begin{subfigure}[b]{0.31\textwidth}
    \centering
    \includegraphics[width=\textwidth]{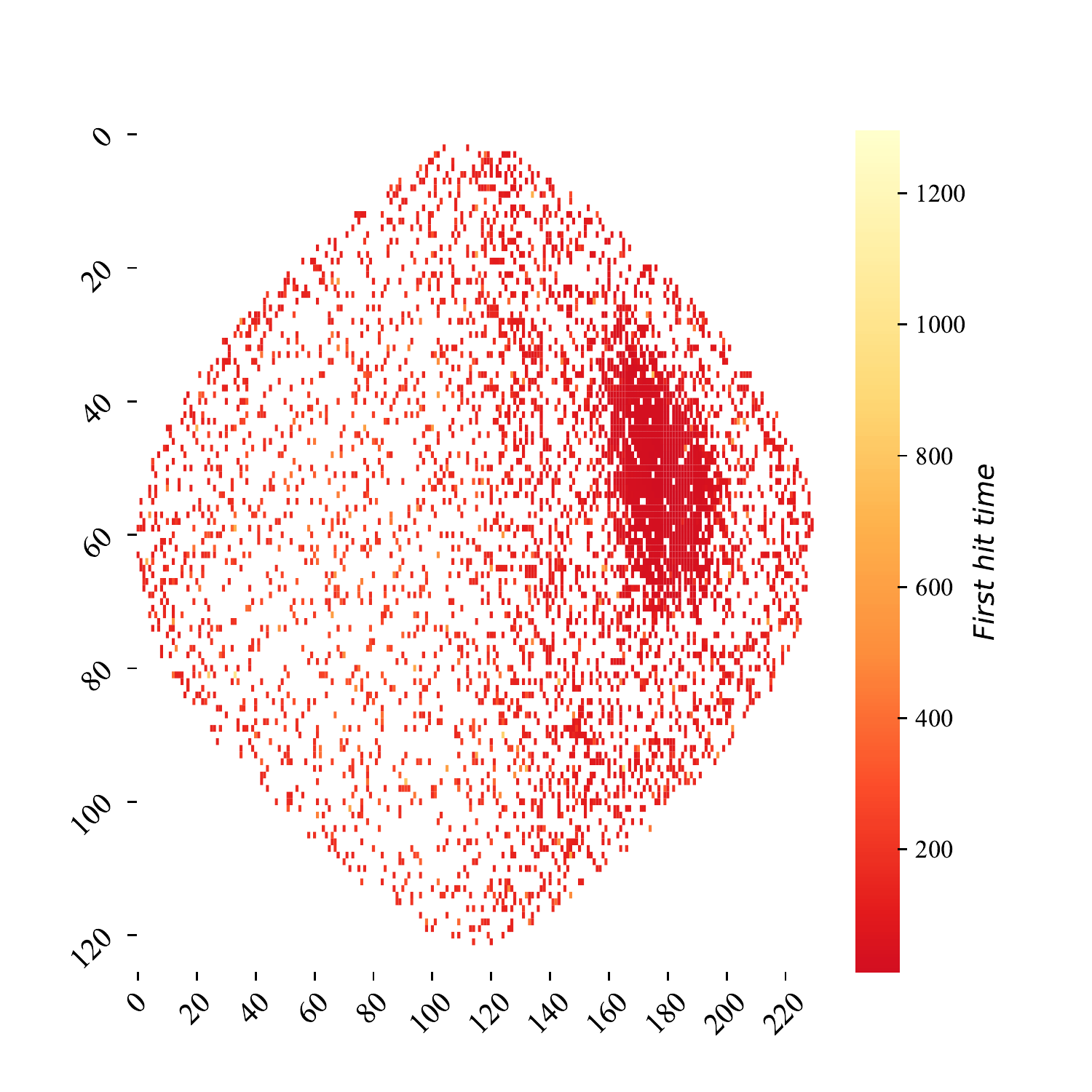}
    \caption{First hit time channel.}
  \end{subfigure}
  \caption{The planar projection method first generates a mapping of the ID of the PMTs and the position of the pixel in the image (a). The charge (b) and first hit time (c) information can be filled in the image according to the mapping.}
  \label{fig:03_2Dimages}
\end{figure}

Using the mapping relationship from the PMT to the pixel position, the hit information obtained by the detector can be converted into the image (see Figure~\ref{fig:03_2Dimages}). The images have two channels: charge and first hit time, which are the inputs of the CNNs.

\subsubsection{VGG-J}
The VGG network is a classic convolutional neural network architecture that still has competitive performance today. It has multiple $3\times3$ kernel-sized filters applied in a chain, one after another. The advantage of using multiple $3\times3$ convolution kernels instead of large convolution kernels is that it has fewer parameters and has better learning ability under the same receptive field. 
There are usually dense layers with thousands of nodes in the last three layers of the original VGG network. The presence of the dense layer makes the network better at fitting, but also greatly increases the amount of parameters. 
\begin{figure}[!ht]
    \centering
  \includegraphics[width=\textwidth]{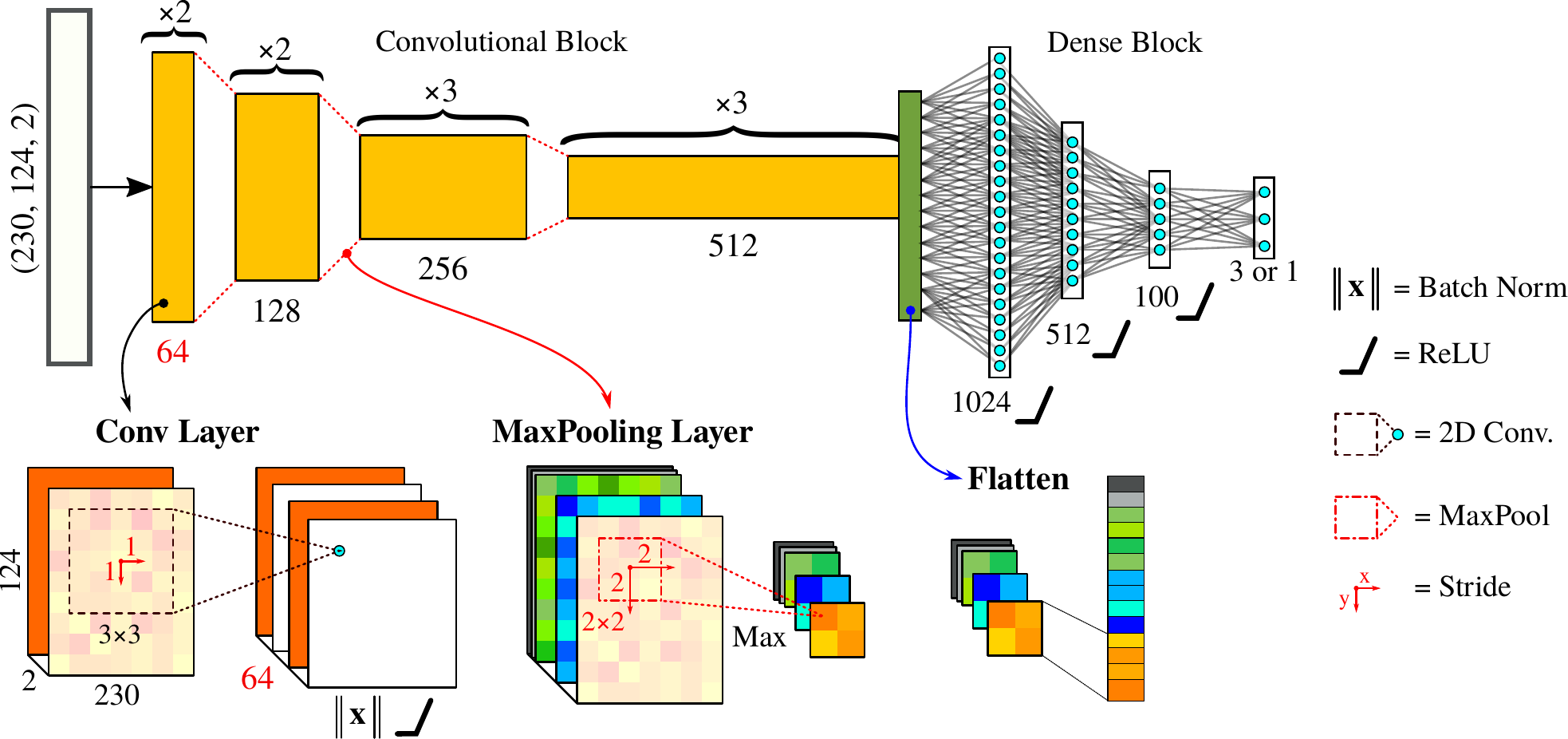}
  \caption{VGG-J network architecture for CNN reconstruction with 17 parameter layers. It is composed of two main blocks: a sequence of $3\times3$ convolutional layers (with maxpooling used for coarsening) and a few dense layers at the end. The last dense layer is used to output the prediction result, which is 1 node for reconstructing energy and 3 nodes for reconstructing vertex coordinates.}
  \label{fig:03-VGG}
  \end{figure}

The VGG-J network we use has 17 layers (see Figure~\ref{fig:03-VGG}). According to the complexity of the energy and vertex reconstruction tasks, we optimize the number of nodes in the final dense layer. The final number of nodes in the last dense layers are: 1024 nodes, 512 nodes and 100 nodes respectively; followed by 1 or 3 nodes, which are used to yield the prediction of energy or vertex.
Compared with the original VGG-16 network that has two layers with 4096 nodes, the amount of parameters in VGG-J network is $26$ million, which has been reduced by $65\%$, while the reconstruction accuracy has remained at the same level.

\subsubsection{ResNet-J}
\begin{figure}[!htb]
    \centering
    \includegraphics[width=0.4\textwidth]{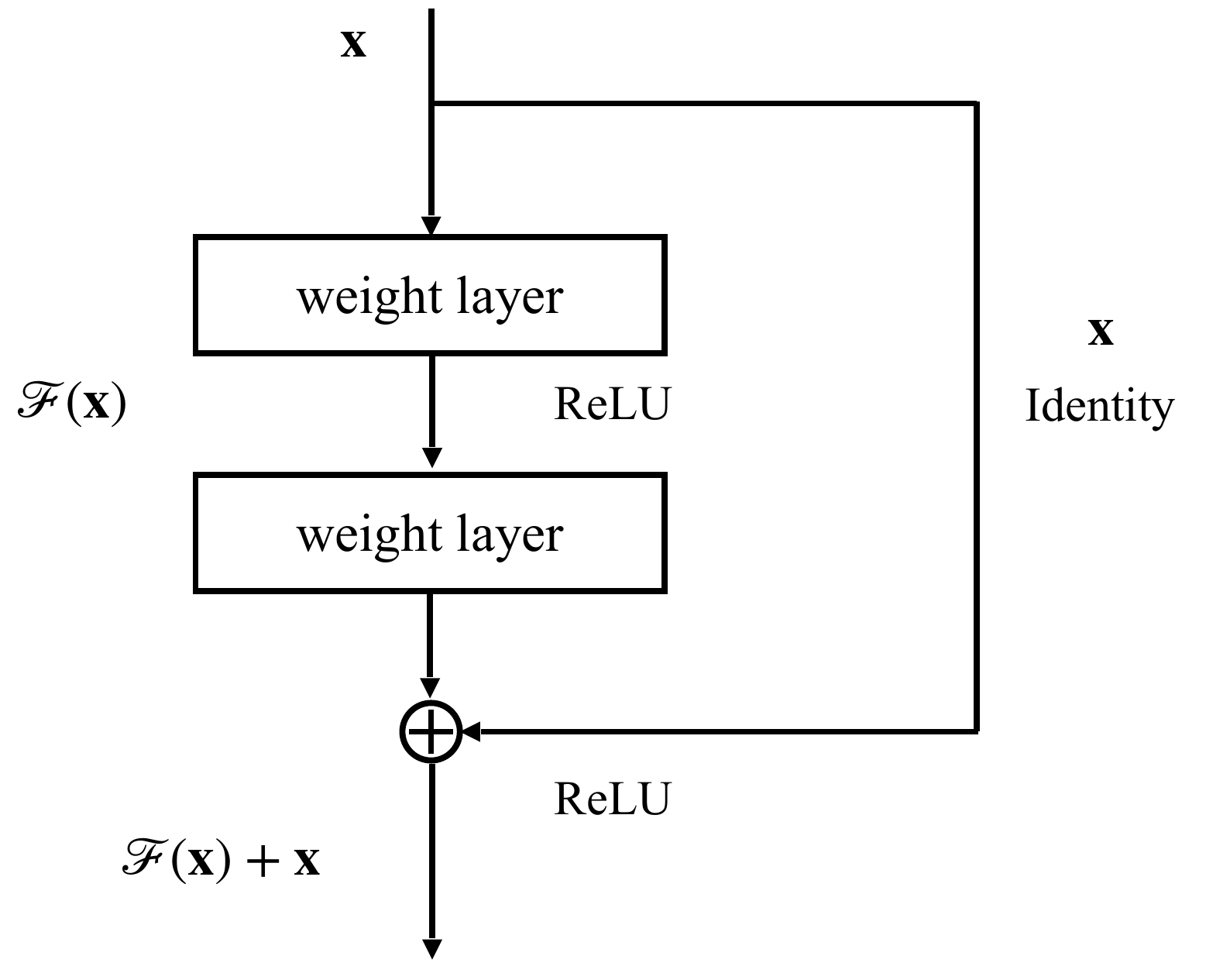}
    \caption{Residual block structure in ResNet network.}
    \label{fig:03_ResNet_block}
  \end{figure}

In order to push the reconstruction performance to the limit, we would like to train a network that has more layers, which may bring better learning ability. However, it will bring undesirable effects if more convolutional layers are added directly, including overfitting, longer training process and slower prediction speed. In order to solve this problem, we use ResNet network architecture~\cite{Resnet_2}. 
The main feature of ResNet is the usage of residual blocks, shown in Figure~\ref{fig:03_ResNet_block},
where $\mathbf{x}$ denotes the input of the block. In a regular NN the block yields the feature mapping $\mathcal{H}(\mathbf{x})$, while the ResNet lets the block fit another feature mapping $\mathcal{F}(\mathbf{x}):=\mathcal{H}(\mathbf{x})-\mathbf{x}$ which is called residual mapping. Therefore, the original mapping is converted into $\mathcal{F}(\mathbf{x})+\mathbf{x}$. It has been discussed that it is easier to optimize the residual mapping than to optimize the original one~\cite{Resnet}. 

\begin{figure}[!htb]
    \centering
    \includegraphics[width=\textwidth]{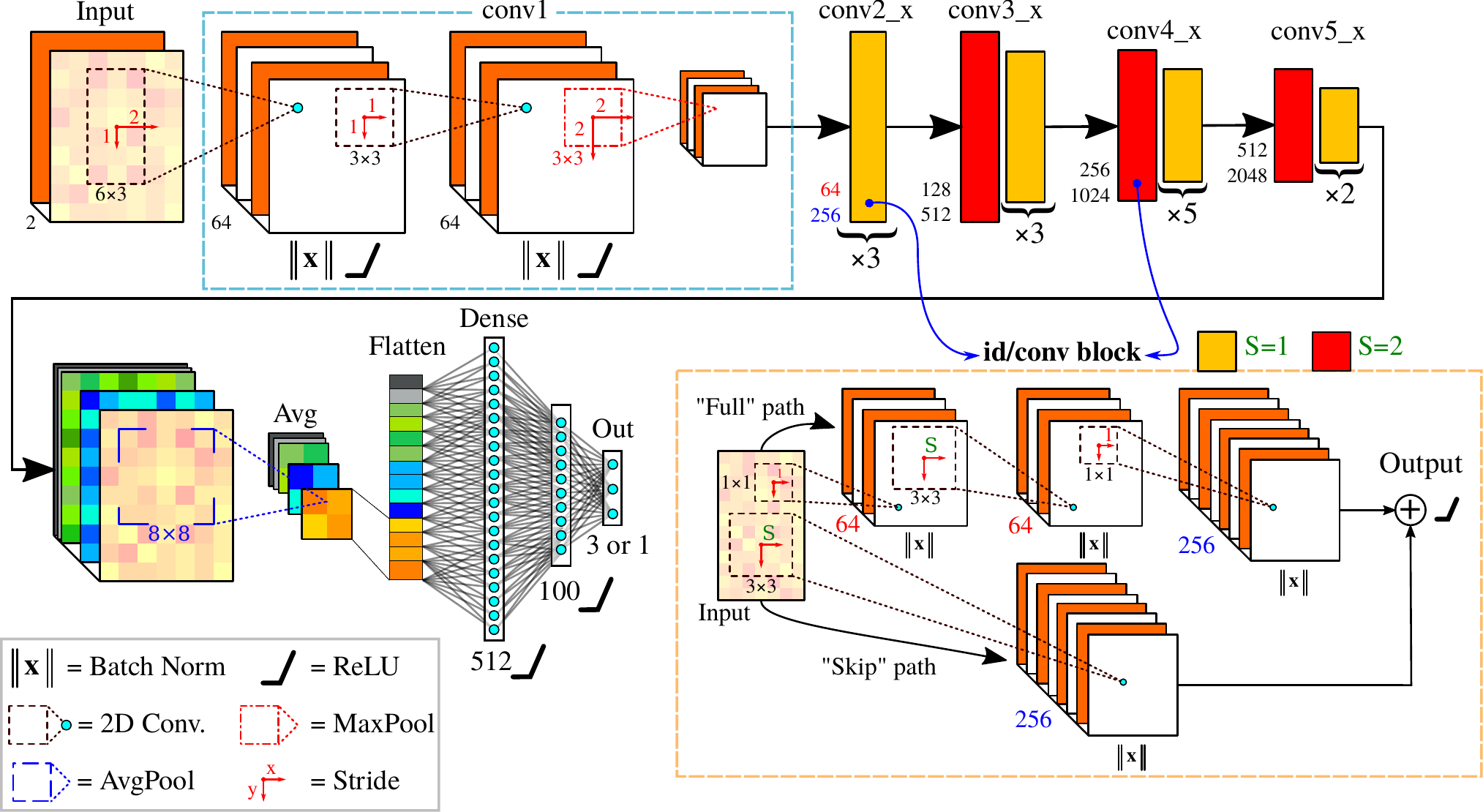}
    \caption{ResNet-J network architecture for CNN reconstruction with $53$ parameters layers. It is composed of a number of residual blocks (see a typical example below) and a couple of dense layers at the end. A residual block is make up of a stack of convolutions. They are $1\times1$, $3\times3$, and $1\times1$ convolutions, where the $1\times1$ layers are responsible for reducing and then increasing (restoring) dimensions and the $3\times3$ layers are responsible for coarsening when it has a stride of $2$ in conv block (red). Next to the main convolutions (Full path), there is a $3\times3$ convolutional layer (Skip path), which has same dimensions and stride so that it can be added to the outputs of stacked layers used for the residual function. Between the convolutions and dense layers, an average pooling layer summarizes all feature got by convolutions, then the dense layers at the end will output the prediction result.}
    \label{fig:03_ResNet}  
  \end{figure}

Compared with the original ResNet50 network, we optimized convolutional layers and the dense layers 
for the reconstruction in JUNO. The final network structure is shown in the Figure~\ref{fig:03_ResNet}
and contains a total of 53~layers with approximately 35~million trainable parameters. In comparison with VGG-J  
it has more layers to enhance the learning ability, see Table~\ref{tab:03_comparsion of VGG and ResNet}. 

\begin{table}[!htb]
    \centering
    \begin{tabular}{lrr}
      \toprule
                                                & VGG-J          & ResNet-J       \\
            \midrule
            Layers                              & 17             & 53             \\
            Parameters                          & \num{26310035} & \num{38352403} \\
            \bottomrule
    \end{tabular}
    \caption{Comparison of VGG-J and ResNet-J architectures.} 
    \label{tab:03_comparsion of VGG and ResNet}
  \end{table}

We used same hyperparameters and training schedule for ResNet-J and VGG-J, see Table~\ref{tab:cnn_hyperparameters}.
It takes about 4~days to train one model on a single V100 GPU.

\begin{table}[htp]
    \centering
    \begin{tabular}{lr}
    \toprule
    \textbf{Parameter} & \textbf{Value}                      \\ 
    \midrule
    Loss               & Mean Squared Error      \\
    Optimizer          & Adam ($\beta_1=0.9$, $\beta_2=0.999$) \\
    Learning rate      & Linearly increasing from 0 to $10^{-3}$ during the first epoch,  \\
                       & then exponential decay to $10^{-8}$.               \\
    Batch size         & 64                                  \\
    N. Epochs          & 15                                  \\ 
    \bottomrule
    \end{tabular}
    \caption{Hyperparameters for VGG-J and ResNet-J.}
    \label{tab:cnn_hyperparameters}
\end{table}

%% file: 03_GNN.tex
\subsection{Spherical Model (GNN-J)} 
\label{sec:03_GNN}

As it was already mentioned in Section~\ref{sec:03_CNN}, the spherical arrangement of PMTs in 
JUNO does not allow to directly use the signal as input for CNNs. One possible workaround is 
to define an arbitrary projection to a Euclidean domain, and then use CNNs as usual, as it was 
done in the previous Section~\ref{ssec:projection}.

However, this comes with a few problems:
\begin{itemize}
    \itemsep0em 
    \item \textbf{Deformation}. Any projection inevitably stretches or shrinks certain areas. So, during convolution, the same filter will capture features coming from spherical regions with different areas and shape, breaking translational invariance and making learning more difficult.
    \item \textbf{Topology}. Distances on the projection are not, in general, proportional to distances on the spherical surface. So, features that are close on the sphere can be far in the 2D projection, meaning that they may not be captured by a local filter.
\end{itemize}


These issues can be avoided by using Graph Neural Networks (GNNs), which generalize CNNs to generic manifolds and remove the need for a projection. 

The main idea is to encode the topology of the input domain in a graph structure, and then properly define convolutions and pooling operations on it. In this work, we adapt the DeepSphere model~\cite{deepsphere}, previously used in cosmology, to the JUNO experiment. The procedure is as follows: 
\begin{parlist}
    \item First, we need to define the graph's nodes which will hold the input samples. A natural choice would be to directly use the PMTs as vertices in 3D space. However, we need also a way to iteratively group neighboring nodes so that their data can be aggregated by the pooling operation. 
    The simplest possibility is to consider a hierarchical partition of the spherical surface, and define nodes in the graph as the regions' centers. In this work, we use the Hierarchical Equal Area isoLatitude Pixelisation (HEALPix) algorithm~\cite{healpix}, which divides the surface into $N_{\mathrm{pix}} = 12 N_{\mathrm{side}}^2$ spherical pixels, all with the same area and centered along rings of equal latitude (see Figure~\ref{fig:healpix}, top). 
    The parameter $N_{\mathrm{side}} \equiv 2^k$ controls the discretization resolution. For the input data, it is set at $N_{\mathrm{side}} = 16$, dividing the detector's surface in $N_{\mathrm{pix}} = 3072$ regions, each containing on average $5.77$ PMTs. Higher values of $N_{\mathrm{side}}$ have been tried (up to $N_{\mathrm{side}}= 64$, at which each pixel contains at most $1$ PMT), but they significantly increase storage and computational requirements, while not improving the reconstruction accuracy.

    A hierarchical discretization means that vertices can be labelled in a nested scheme (see Figure~\ref{fig:healpix}, bottom), which makes pooling operations very efficient.

\begin{figure}[htp]
    \centering
    \includegraphics[width=0.9\textwidth]{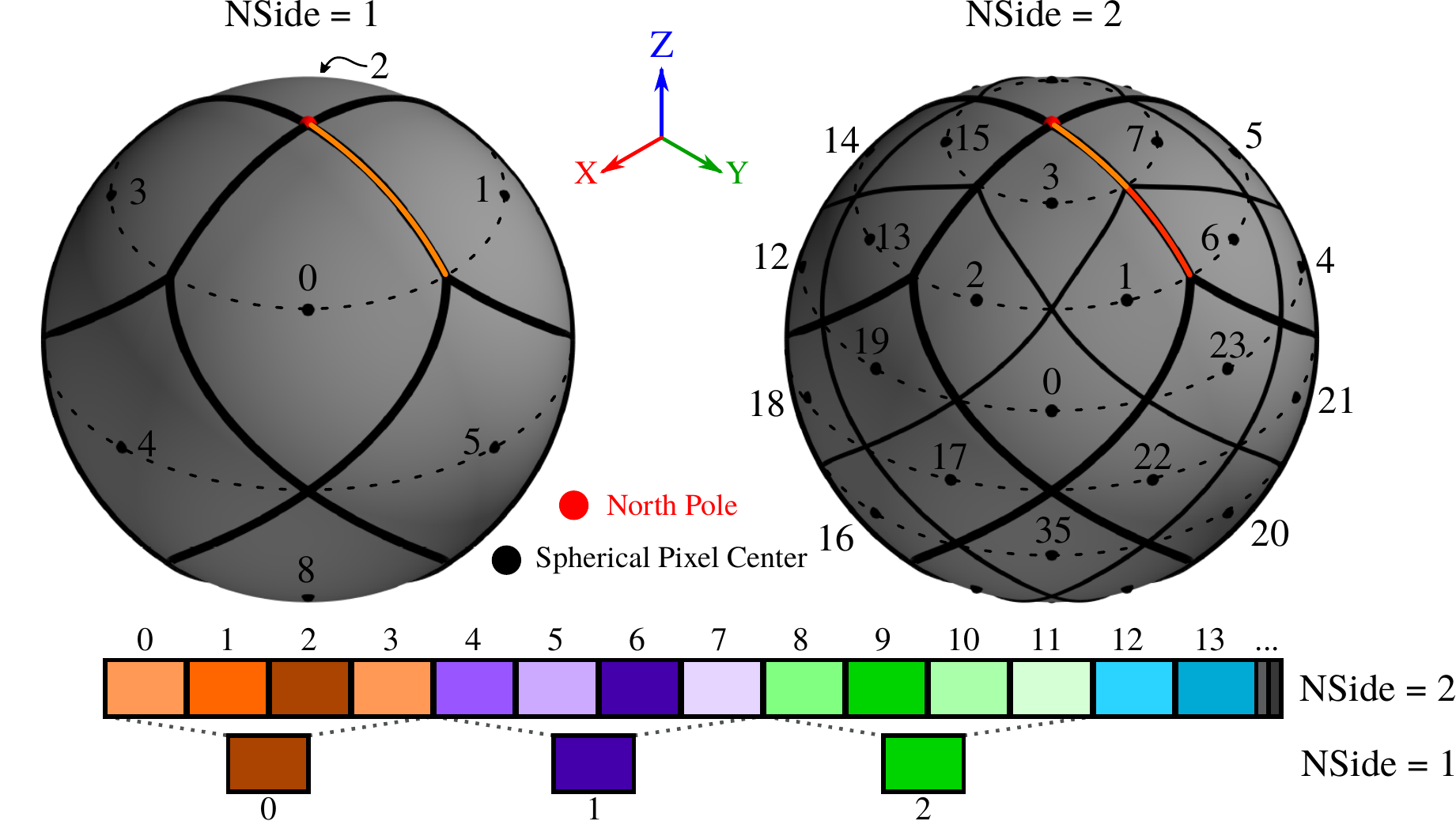}
    \caption{The HEALPix algorithm starts by dividing the spherical surface into $12$ regions: $4$ around each pole, and $4$ at the equator. Then, resolution can be increased by iteratively dividing each spherical pixel into $4$ sub-pixels of equal area, so that any edge of the original $12$ regions is split into $N_{\mathrm{side}}$ parts. Pixels are labelled in a nested scheme, so that subpixels belonging to the same region have consecutive indices. In this way, a pooling operation on the discretized sphere is as efficient as the usual 1D pooling. As an example, consider the bottom picture. Pixels belonging to the same region have similar colors, with their brightness representing different values. The pooling operation consists of aggregating all pixels in the same region into a single value, which in this example is given by the darkest color.}
    \label{fig:healpix}
\end{figure}

    \item We construct a simple, undirected graph $\mathcal{G} = (\mathcal{V}, \mathcal{E}, \mathrm{W})$ encoding the discretization structure. In this notation, $\mathcal{V} = \{\bm{v_i}\}_{i=1,\dots,N_{\mathrm{pix}}}$ is the set of $N_{\mathrm{pix}}$ vertices, with $\bm{v_i} \in \mathbb{R}^3$ being the center in 3D space of the $i$-th spherical pixel. Then, 
    $\mathcal{E} \subset \mathcal{V}\times \mathcal{V}$ is the set of active links between vertices, and $\mathrm{W} \in \mathbb{R}^{N_{\mathrm{pix}} \times N_{\mathrm{pix}}}$ is the positive symmetric weighted adjacency matrix, such that $\mathrm{W}_{ij}$ is the weight of the connection from node $i$ to $j$, representing their \q{closeness}, with $\mathrm{W}_{ij} > 0$ if and only if $(\bm{v_i},\bm{v_j}) \in \mathcal{E}$.

    The choice of the connection weights is discretionary. For simplicity, we adopted the same convention used by the DeepSphere model~\cite{deepsphere}, for which only neighboring pixels $i$, $j$ have non-zero weights $\mathrm{W}_{ij}$ defined by a Gaussian function:
    \begin{align}
        \mathrm{W}_{ij} = \exp \left( - \frac{\norm{\bm{v_i} - \bm{v_j}}^2_2}{2 \overline{d^2}} \right), \qquad \overline{d^2} = \frac{1}{|\mathcal{E}|} \quad \sum_{\mathclap{(\bm{v_i},\bm{v_j}) \in \mathcal{E}}} \> \norm{\bm{v_i} - \bm{v_j}}_2^2,
        \label{eqn:adjacency-weights}
    \end{align}
    where $\norm{\bm{x}}_2 \equiv \sqrt{x_1^2 + \dots + x_n^2}$ denotes the Euclidean norm, and $|\mathcal{E}|$ is the number of elements in the set $\mathcal{E}$, i.e. the number of links in $\mathcal{G}$.     
    The idea of (\ref{eqn:adjacency-weights}) is that the connection weight $\mathrm{W}_{ij}$ between two nodes is higher if they are closer, and decays quickly with their (squared) distance. Note that, since we are only connecting neighboring nodes, their distance on the spherical surface can be locally approximated by the Euclidean norm. Then, the average $\overline{d^2}$ of the squared distances is used to normalize the argument of the exponential.

    With this choice, the nodes are only locally connected, meaning that $\mathrm{W}$ is sparse, i.e. contains mostly zeroes, and so computations may be optimized. To construct a GNN, the only requirement for the $\mathrm{W}_{ij}$ is to encode a connected graph, i.e. such that there is a set of edges with non-zero $\mathrm{W}_{ij}$ linking any two nodes $i$ and $j$.  However, we did not investigate different choices for (\ref{eqn:adjacency-weights}).
    \item An input sample $\bm{x} \in \mathbb{R}^{N_{\mathrm{pix}}\times F}$ is a signal on $\mathcal{G}$, i.e. a function mapping each node to a vector of $F$ features. 
    
    In this work $F=2$, and we consider, for each PMT, its charge (i.e. the number of PE hits) and its first hit time, relative to the event's origin. 
    If a PMT receives no hits, we assign a discretionary first hit time of $\SI{1024}{\nano\s}$, denoting that it is hit \q{at infinity}. 

    Since spherical pixels contain more than one PMT, we need to aggregate data from several PMTs to form the feature vector $\bm{x_i} \in \mathbb{R}^F$ of the $i$-th spherical pixel. So, for every spherical pixel, we sum the charges of all the PMTs inside it, and take the minimum of their first hit times.
    \item Before the training, we normalize each channel (charge and first hit) in the training dataset to $0$ mean and unit standard deviation. In this way, all the features have the same order of magnitude, which is necessary for the model to converge.
    \item Convolutions on $\mathcal{G}$ can be defined in many ways. In this work we use Chebyshev Convolutional Layers~\cite{spectralfilters}, which use the spectral domain of the graph to define filters.  
    \item The model is implemented using the Spektral library~\cite{spektral} and Tensorflow~2.2~\cite{tensorflow} with \texttt{tf.Keras}.
    
    The architecture, further referred to as GNN-J (see Figure~\ref{fig:deepsphere_architecture}), is inspired by that of VGG-16, with some minor changes in the number of filters/layers which resulted in a small ($\sim 5\%$) improvement in validation accuracy. All the model's hyperparameters are summarized in Table~\ref{tab:deepsphere_hyperparameters}. They were found by a manual trial and error over a small set of alternatives. In fact, since training takes $\sim \SI{22}{\hour}$ on a single V100 GPU, it was not feasible to perform a more comprehensive automated search.

\begin{figure}[htp]
    \centering
    \includegraphics[width=\textwidth]{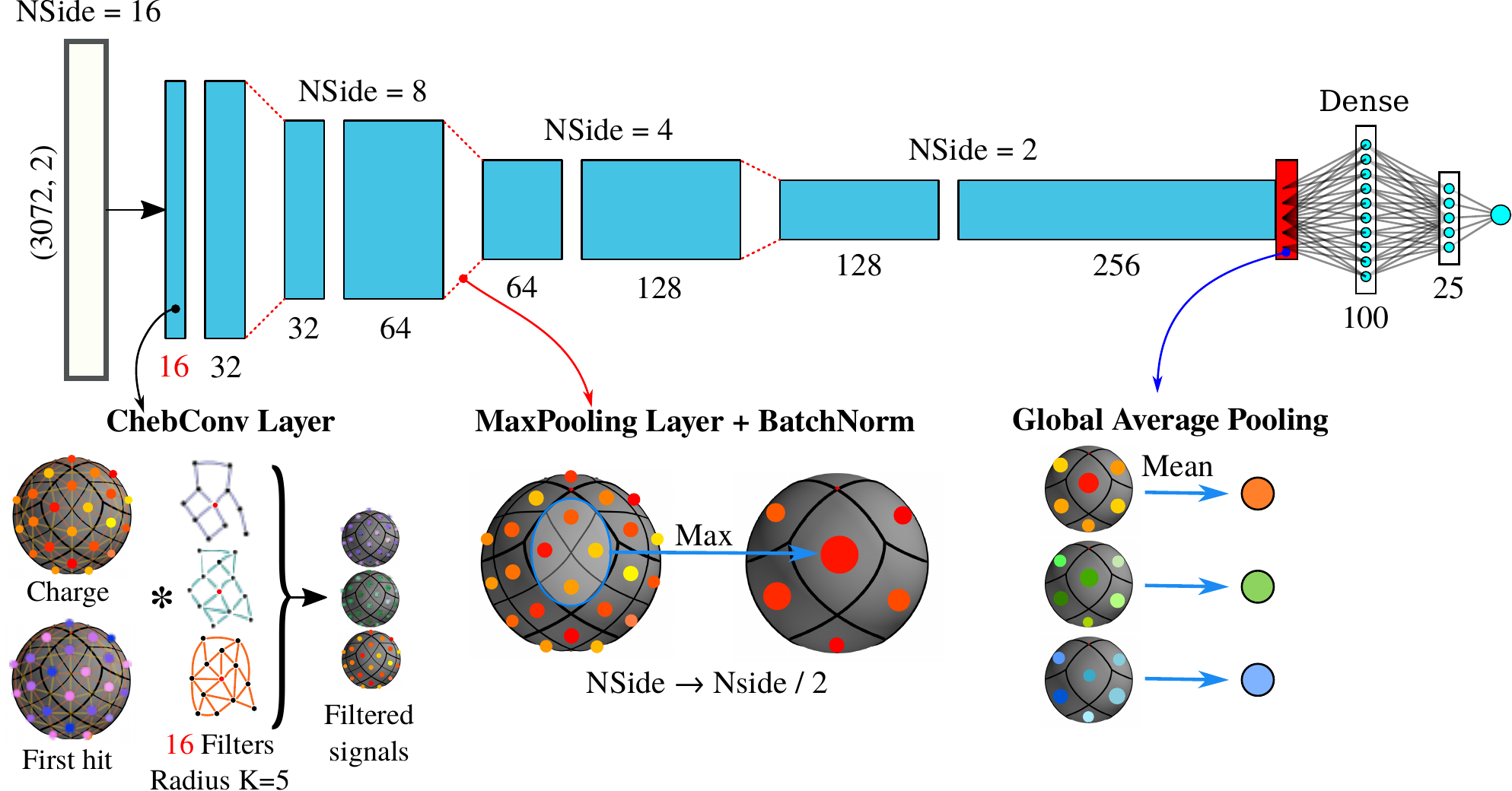} 
    \caption{Architecture for the GNN-J model. It is composed of two main blocks: a sequence of Chebyshev convolutional layers (with maxpooling used for coarsening) and a couple of dense layers at the end. Between the two, a global average pooling layer computes averages for each filter, leading to a certain degree of rotational invariance. Graph convolutions happen at the spectral domain, and involve filters that are localized, i.e. with a finite (graph) radius $K = 5$. Their topology is parametrized by the coefficients of a $K$-order Chebyshev polynomial, which are part of the model's learnable parameters. \label{fig:deepsphere_architecture}} 
\end{figure} 

\begin{table}[htp]
    \centering
    \begin{tabular}{lr}
    \toprule
    \textbf{Parameter} & \textbf{Value}                      \\ \midrule
    Loss               & Mean Absolute Percentage Error      \\
    Optimizer          & Adam ($\beta_1=0.8$, $\beta_2=0.9$) \\
    Learning rate      & Fixed at $0.001$ for $N_{\mathrm{epoch} } < 3$, then exponential decay at rate $-0.1$.                      \\
    Batch size         & 64                                  \\
    N. Epochs          & 10                                  \\ \bottomrule
    \end{tabular}
    \caption{Hyperparameters for GNN-J.}
    \label{tab:deepsphere_hyperparameters}
\end{table}

    As a final detail, we note that using a relative loss, such as the Mean Absolute Percentage Error (MAPE), works best for the task of energy reconstruction, improving resolution and bias at low energies. However, it also makes training more unstable: sometimes a bad initialization results in an initial loss of $100\%$, which does not improve over time. In these cases, weights need to be re-initialized, and the training restarted.
\end{parlist}

%% file: 04_definitions.tex
In the following sections we will present the performance of the studied methods
(BDT, DNN, ResNet-J, VGG-J and GNN-J) for the reconstruction of primary vertex and
energy.
Only one of the models, GNN-J is used exclusively for the energy reconstruction.
The task of the vertex reconstruction requires the time information of each PMT taken into account.
Since the current implementation of GNN-J aggregates data to a some degree even at the
input level, it is not suitable for the task.
Ways to overcome this limitation will be discussed in Section~\ref{sec:discussion}.

Before comparing the results, we also present an overview of performance parameters and outline their
expected behavior.

\subsection{Definition of the performance parameters}
\label{ssec:definitions}

\begin{figure}[!htb]
    \centering
    \includegraphics[width=\textwidth]{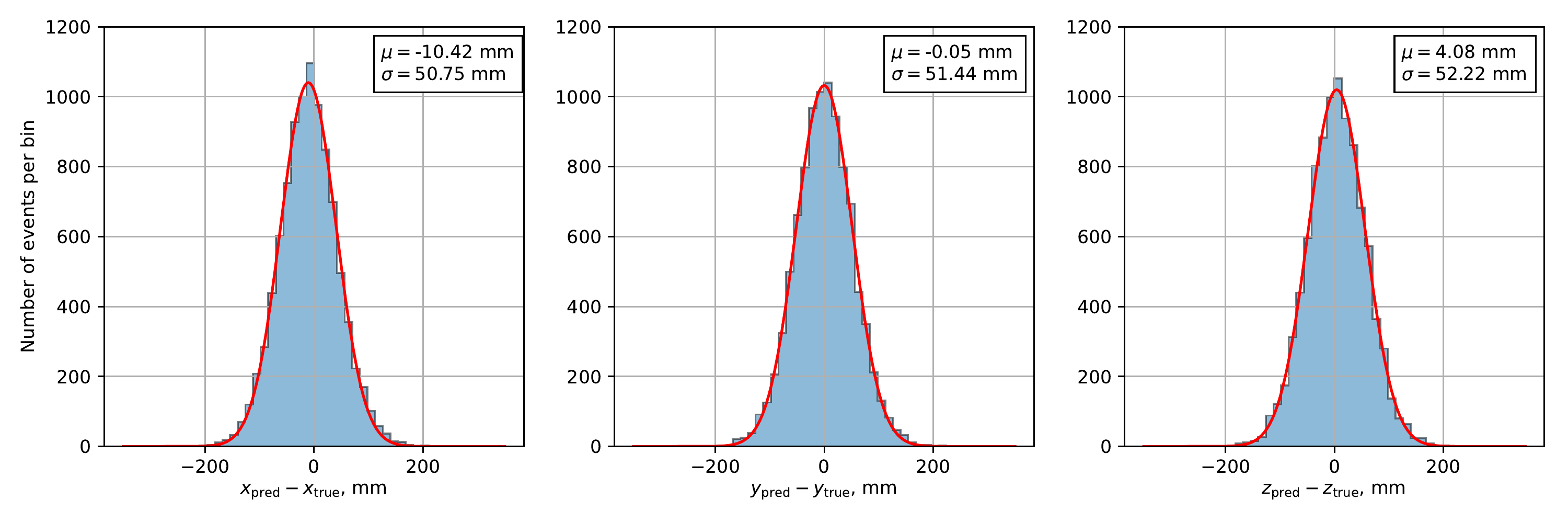}{\centering}
    \caption{An example of Gaussian fit for a spatial variable in each direction for
$E_{\text{vis}}$ = \SI{4.022}{\MeV}, used to extract the bias and the resolution.
The predictions are produced with VGG-J.}
    \label{fig:gaus_fit_v}
\end{figure}

In order to evaluate the performance of the trained models, both the neural networks and the decision
trees, we study two characteristics: resolution and bias. They are defined
by a Gaussian fit, as shown in Figure~\ref{fig:gaus_fit_v}. The mean value of the
best fit Gaussian corresponds to the reconstruction bias and represents the
systematic shift introduced by the reconstruction, which potentially may be compensated.
The value of the $\sigma$ of the Gaussian corresponds to the reconstruction resolution.
This approach is used for both the vertex and the energy reconstruction. The uncertainties of
the fit values are shown on the plots with vertex and energy resolution by error bars.

Bias and resolution are studied as a function of two variables. The first one is visible energy. It is
a combination $E_{\text{vis}}=E_e+m_e=E_{\rm kin}+1.022$~MeV of the total
positron energy and the electron mass, which appears due to positron-electron annihilation.
The light collection and the number of triggered PMTs grow with the energy, which makes
the reconstruction more precise.

The detector is symmetric versus rotation; therefore, the main difference in reconstruction
arises from a distance between the detector center and the vertex --- its radial position, which
is used as a second variable. Events in the center of the detector produce a more symmetric response.
The events on the edges of the detector are affected by the light attenuation in the LS, effects of
the light scattering and re-emission and, near the edge, by the total internal reflection in the
acrylic sphere. The results are sampled versus $r^3$, since cubic sampling produces equal volume
spherical layers and provides equal statistics samples.

The performance of the vertex reconstruction is studied as a function of both visible energy and
radial position. It is reported in absolute values in \si{\milli\m}.

It is worth noting that while the angular resolution is high, the bias and resolution of the radial
component do not directly correspond to the Cartesian distance between the true and reconstructed
vertex.

The performance of the energy reconstruction is studied as a function of visible energy and is
reported as a ratio to the visible energy in percents. It could in principle be interesting
to study energy resolution as a function of number of photo-electrons, because to the first order the resolution
is defined as $1/\sqrt{n_{\rm p.e.}}$.
However, for the large JUNO detector $n_{\rm p.e.}$ depends on the event position in the detector. By this reason we do
not present the results as function of $n_{\rm p.e.}$.

\begin{figure}[!htb]
    \centering
    \includegraphics[width=0.48\textwidth]{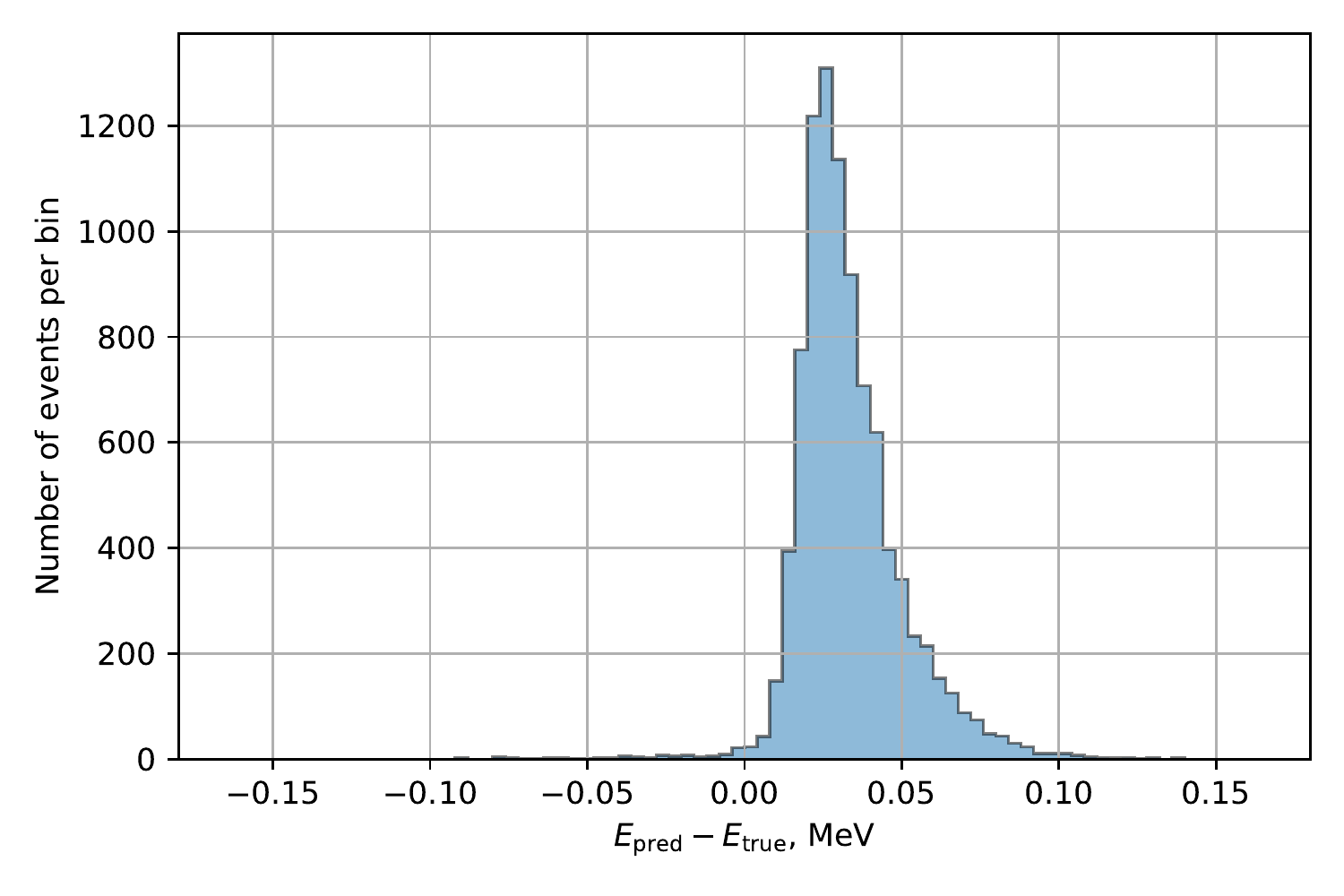}
    \includegraphics[width=0.48\textwidth]{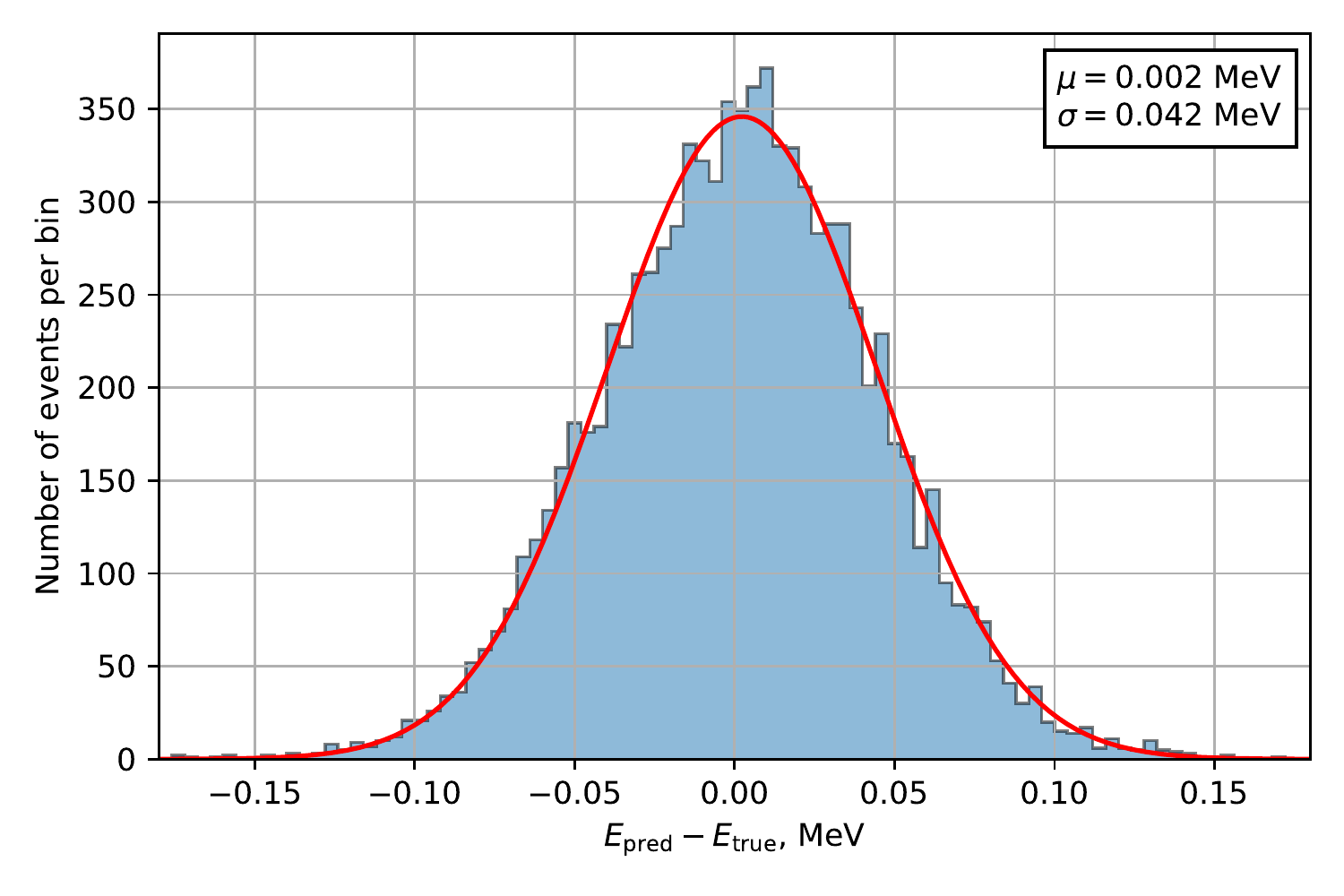}
    \caption{An example of energy prediction distributions for $E_{\text{vis}}$ = \SI{1.022}{\MeV} (left) and
    $E_{\text{vis}}$ = \SI{2.022}{\MeV}  (right). The latter one is fit with Gaussian function.
    The predictions are produced with BDT.}
    \label{fig:gaus_fit_e}
\end{figure}

It has to be noted here that
the ML models learn that the event energy belongs to the range of
the training dataset (\SIrange{0}{10}{\MeV}) and never happens
outside, therefore the distribution of the reconstructed energy at the edges of the dataset becomes
asymmetric, as shown in Figure~\ref{fig:gaus_fit_e}.
In order to simplify the following considerations, we do not analyze the points on the edges of the
dataset. Instead, we only consider points from \SI{0.1}{MeV} to \SI{9.0}{MeV}, for which
the prediction distributions are well fit by Gaussian.
Since the edge values are outside the region of interest of physics, which has a range of
\SIrange{0.5}{9.0}{\MeV}, no important information is lost by the truncation.

%% file: 04_vertex.tex
\subsection{Vertex Reconstruction}
\label{sec:vertex_results}

\begin{figure}[!htb]
    \centering
    \begin{subfigure}{0.5\linewidth}
      \includegraphics[width=\textwidth]{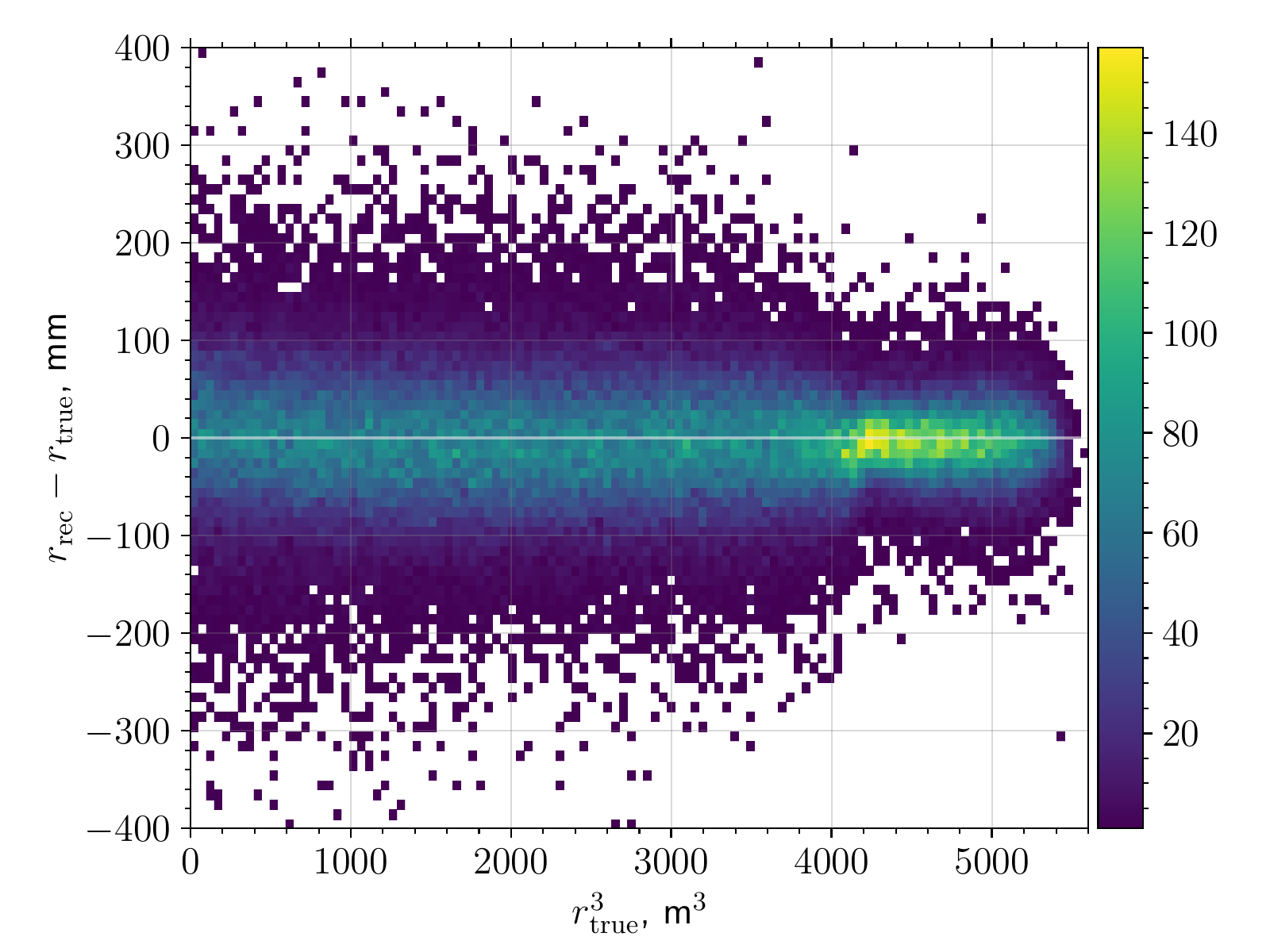}
    \end{subfigure}%
    \begin{subfigure}{0.5\linewidth}
      \includegraphics[width=\textwidth]{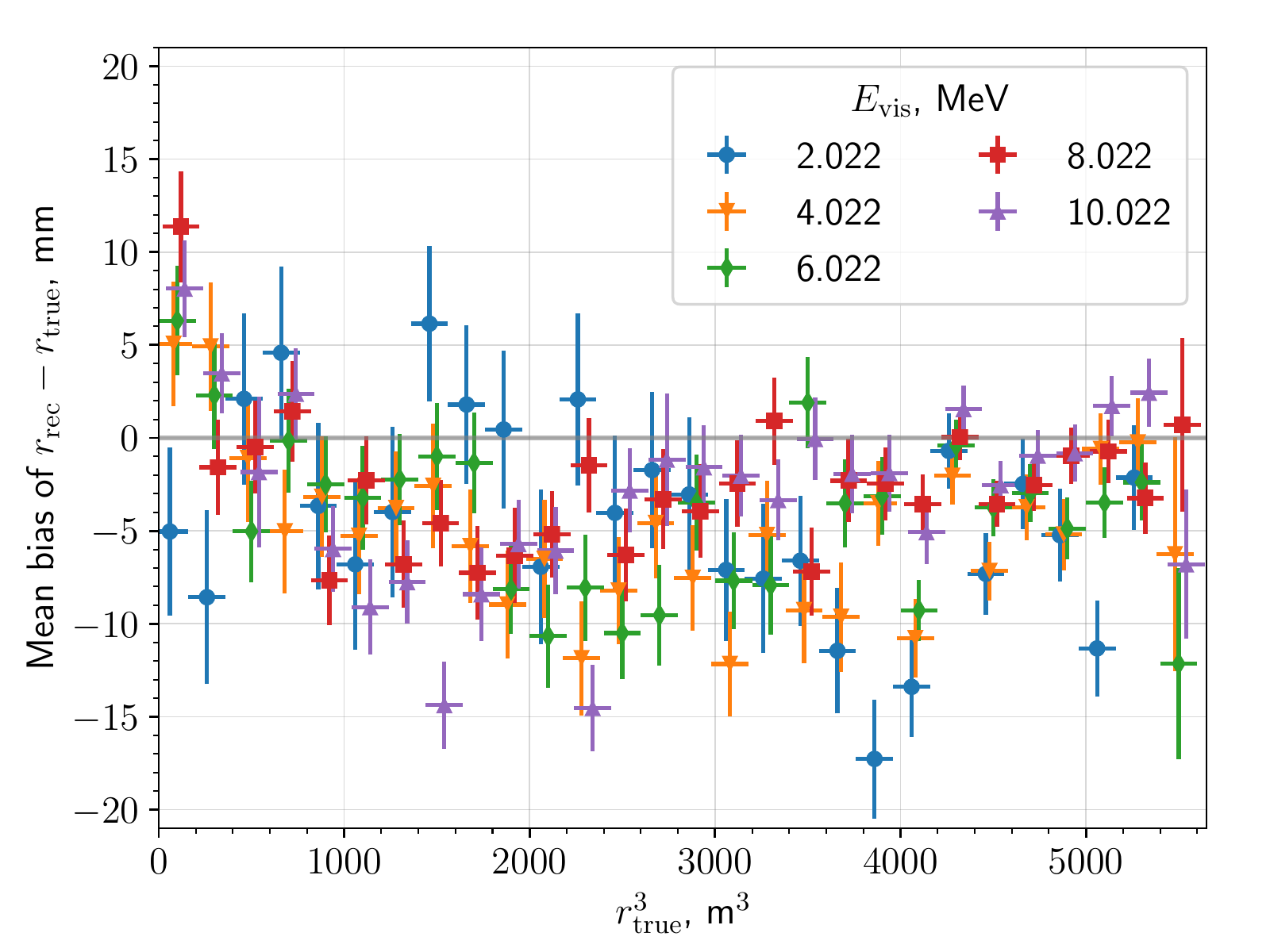}
    \end{subfigure}
    \caption{Heatmap of $r_{\rm rec}-r_{\rm true}$ versus $r^{3}_{\rm true}$ for
      all the testing data samples (left). Mean bias versus $r^{3}_{\rm true}$
      for different energies (right). The results are obtained with VGG-J
      taking TTS and DN into account. Error bars correspond to the standard error of the mean. 
      The plot on the right panel is offset along X-axis within $\pm\SI{40}{\m^3}$ for better readability.
      }
    \label{fig:bias}			
\end{figure}    

	The current VGG-J result shows that the absolute value of bias is less than \SI{15}{\milli\m} 
in the whole detector when taking TTS and DN into account, see Figures~\ref{fig:bias} (left) and 
it is not energy-dependent, see Figure~\ref{fig:bias} (right).

\begin{figure}[!htb]
    \centering
    \includegraphics[width=0.49\textwidth]{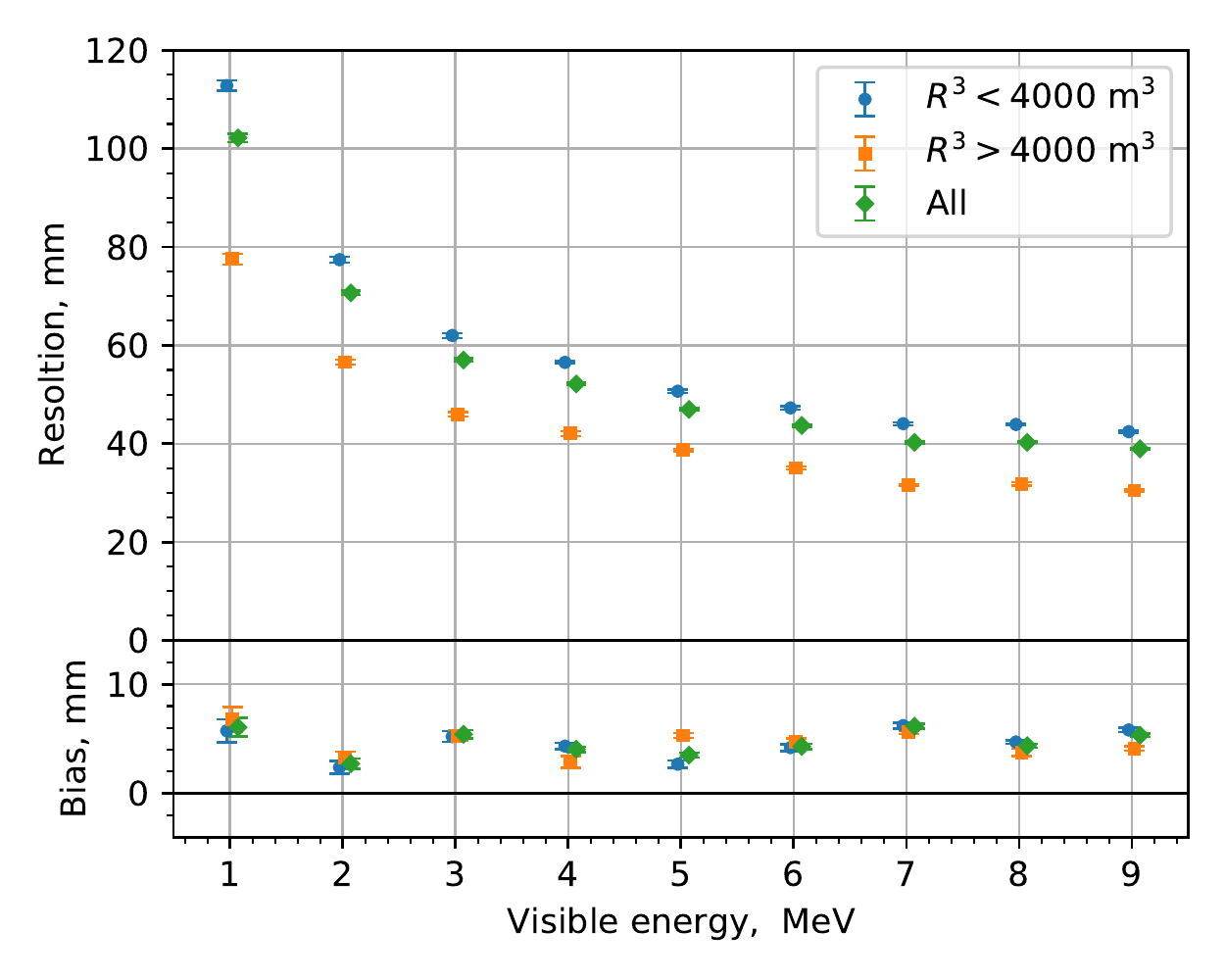}
    \includegraphics[width=0.49\textwidth]{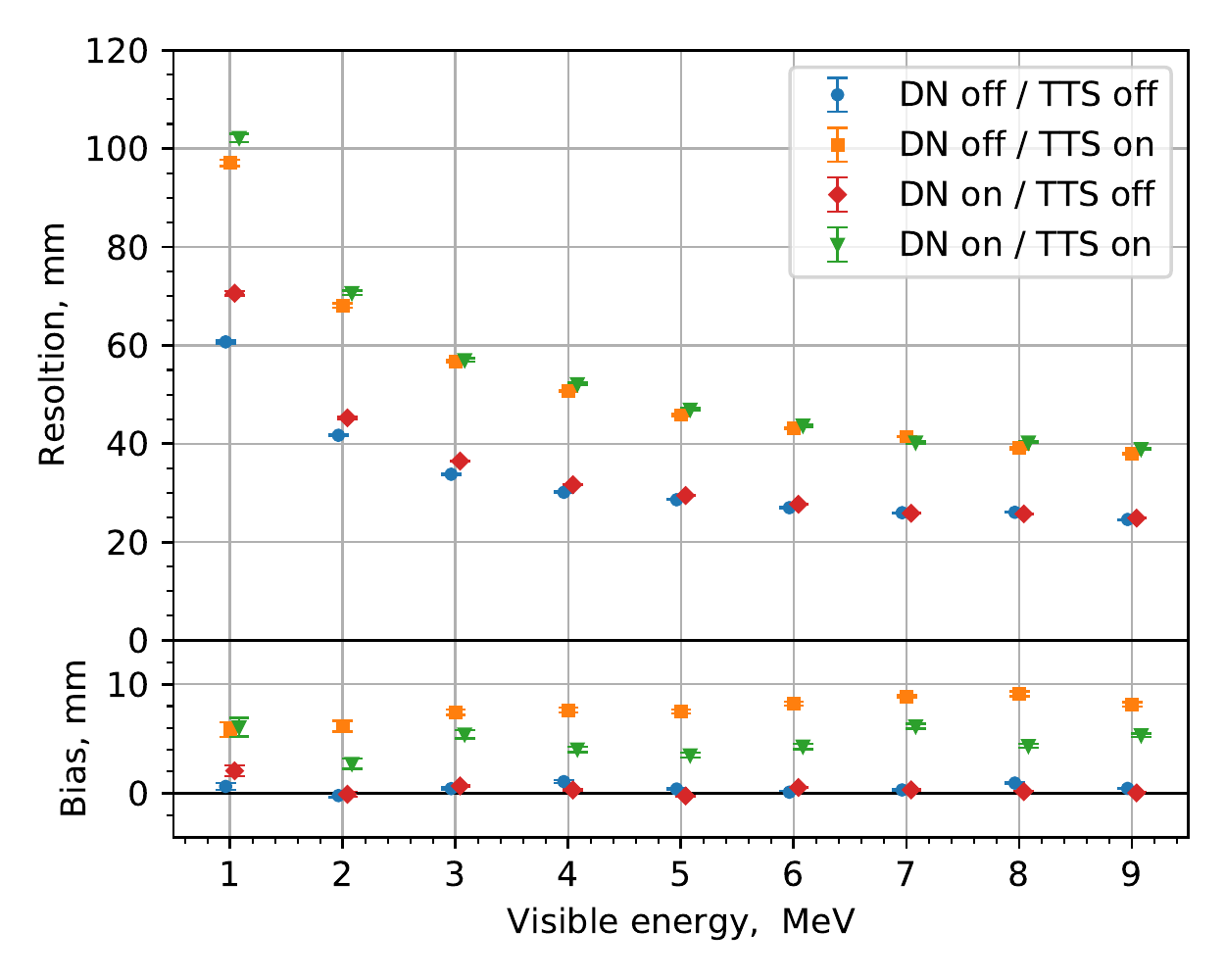}
    \caption{Resolution of the vertex $z$-coordinate obtained with VGG-J taking TTS and DN 
      into account for different part of the detector (left) and the resolution of the whole 
      detector for different TTS/DN options (right). The plots are offset along X-axis within 
      $\pm\SI{0.06}{\MeV}$ for better readability.}				
    \label{fig:resdiff}			
\end{figure}

	From Figure~\ref{fig:bias} (left), it is clear that the resolution is much better in the 
border region of the detector ($R^{3} >$ \SI{4000}{\m\cubed}), than the inner region of the 
detector ($R^{3} <$ \SI{4000}{\m\cubed}). The reason is that TTS of the NVVT PMTs is 
relatively large, and so the charge information could provide more strict constraint on the vertex 
position than the first hit time information when the vertex is close to the PMTs. Therefore, 
we separate the JUNO detector into two parts to evaluate the resolution, as shown in 
Figure~\ref{fig:resdiff} (left). The vertex resolution is about \SI{8.0}{\centi\m} at 
\SI{1}{\MeV} and decreases to \SI{3.1}{\centi\m} at \SI{9}{\MeV} for the border region of the 
detector, while it is about \SI{11.5}{\centi\m} at \SI{1}{\MeV} and decreases to 
\SI{4.2}{\centi\m} at \SI{9}{\MeV} for the inner region of the detector.

	The influence of TTS and DN on the vertex reconstruction is shown in Figure~\ref{fig:resdiff} (right). The results show that TTS is the main influencing factor leading to the worsening of vertex resolution, while the DN impact is negligible, which is reasonable because the time information, that is affected by TTS, is exploited by  vertex reconstruction the most.

	\begin{figure}[!htb]
		\centering
        \includegraphics[width=0.6\textwidth]{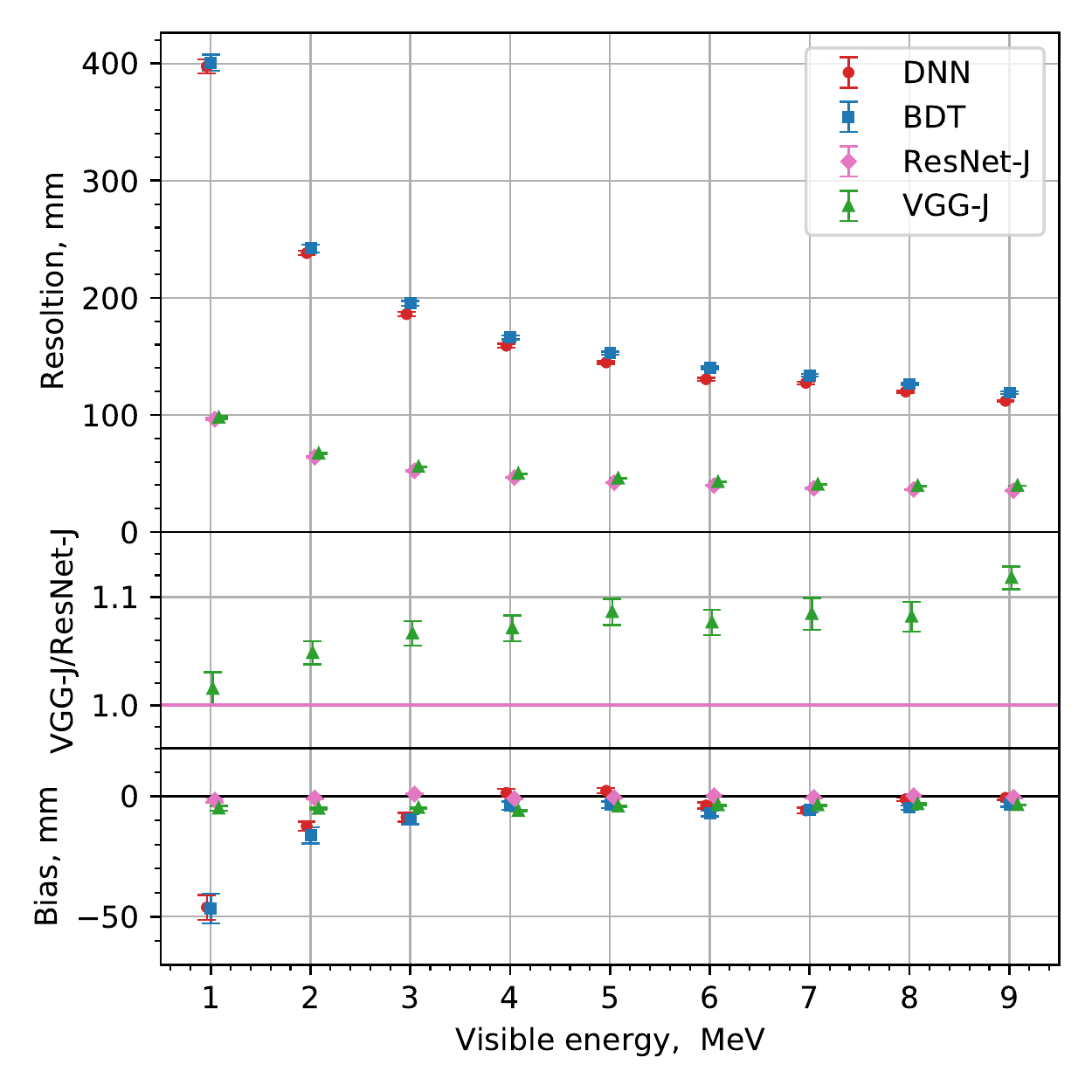}
		\caption{Resolution of the vertex radial component of DNN, BDT, ResNet-J and VGG-J models (upper panel),
                 ratio between VGG-J and ResNet-J resolution (middle panel),
                 and bias for all the four models (lower panel). 
                 The plots are offset along X-axis within $\pm\SI{0.06}{\MeV}$ for better readability.}
		\label{fig:vgg16vsres50}			
	\end{figure}    

  In a more realistic case that takes TTS and DN into account, the performance comparison of the four models is shown in Figure~\ref{fig:vgg16vsres50}. 
The PMT-wise measured information, containing both the hit time and the number of photo-electrons, is of great help to the vertex reconstruction because of its details, which also offers a decisive advantage for algorithms that use it. 
So it is not a surprise that VGG-J and ResNet-J can provide sufficient resolution, while BDT and DNN with aggregated information can not.

Figure~\ref{fig:vgg16vsres50} (middle panel) shows the ratio of resolution between VGG-J and ResNet-J. Generally, ResNet-J performs slightly better. 
ResNet-J and VGG-J introduce small bias,  and for the other models, the bias is within 25 mm (except one very first point), as shown in Figure~\ref{fig:vgg16vsres50} (lower panel). The conclusion is similar for other TTS/DN options.
%

%% file: 04_energy.tex
\subsection{Energy Reconstruction}
\label{sec:energy_results}

To estimate energy resolution and bias we analyze the distribution of 
$(E_{\rm pred}-E_{\rm true})/E_{\rm true}$.
Prior to analysis, a small fraction of outliers is 
trimmed away by excluding the values deviating by more than three standard deviations from the 
mean value. This procedure allows getting rid of events where a part of the energy is carried out by 
gammas leaking the detector. The fraction of such events lies in the range from 0.3\% to 0.5\%.

\begin{figure}[!htb]
    \centering
    \includegraphics[width=0.6\textwidth]{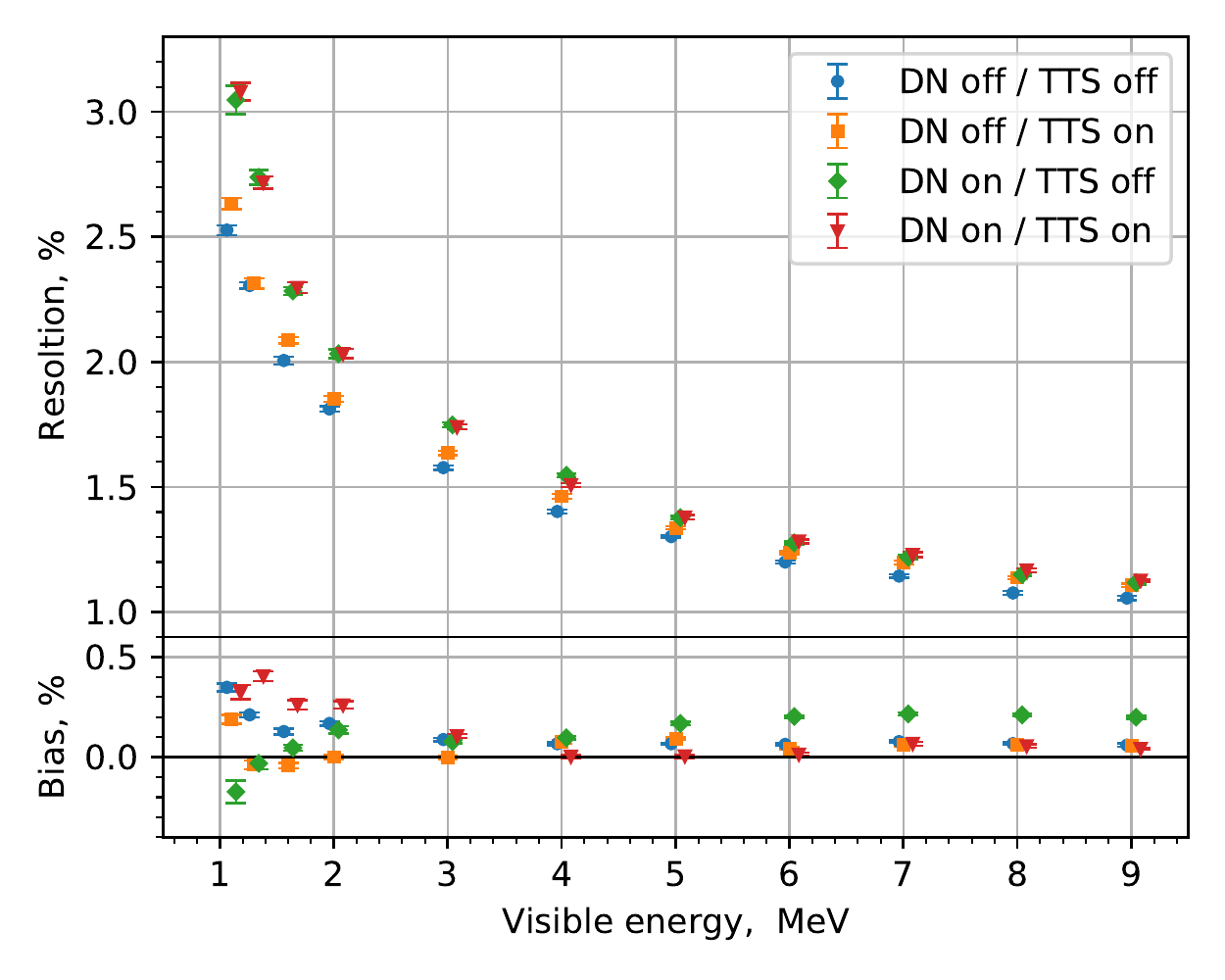}
    \caption{Energy reconstruction performance: resolution (upper panel) and bias (lower panel) 
    obtained with DNN. The plots are offset along X-axis within $\pm\SI{0.06}{\MeV}$ for better readability. 
    Note that the first point corresponds to 1.122 MeV (see explanation in Section~\ref{ssec:definitions}).}
    \label{fig:ene_models}
\end{figure}

Figure~\ref{fig:ene_models} shows how the performance of DNN degrades when taking TTS and DN into account. Generally TTS does not make much impact on the energy resolution, while one can see that DN significantly affects it. Bias is kept within 0.5\% and 
increases with inclusion of the electronics effects. The other models share a similar behavior.

\begin{figure}
    \centering
    \includegraphics[width=0.6\textwidth]{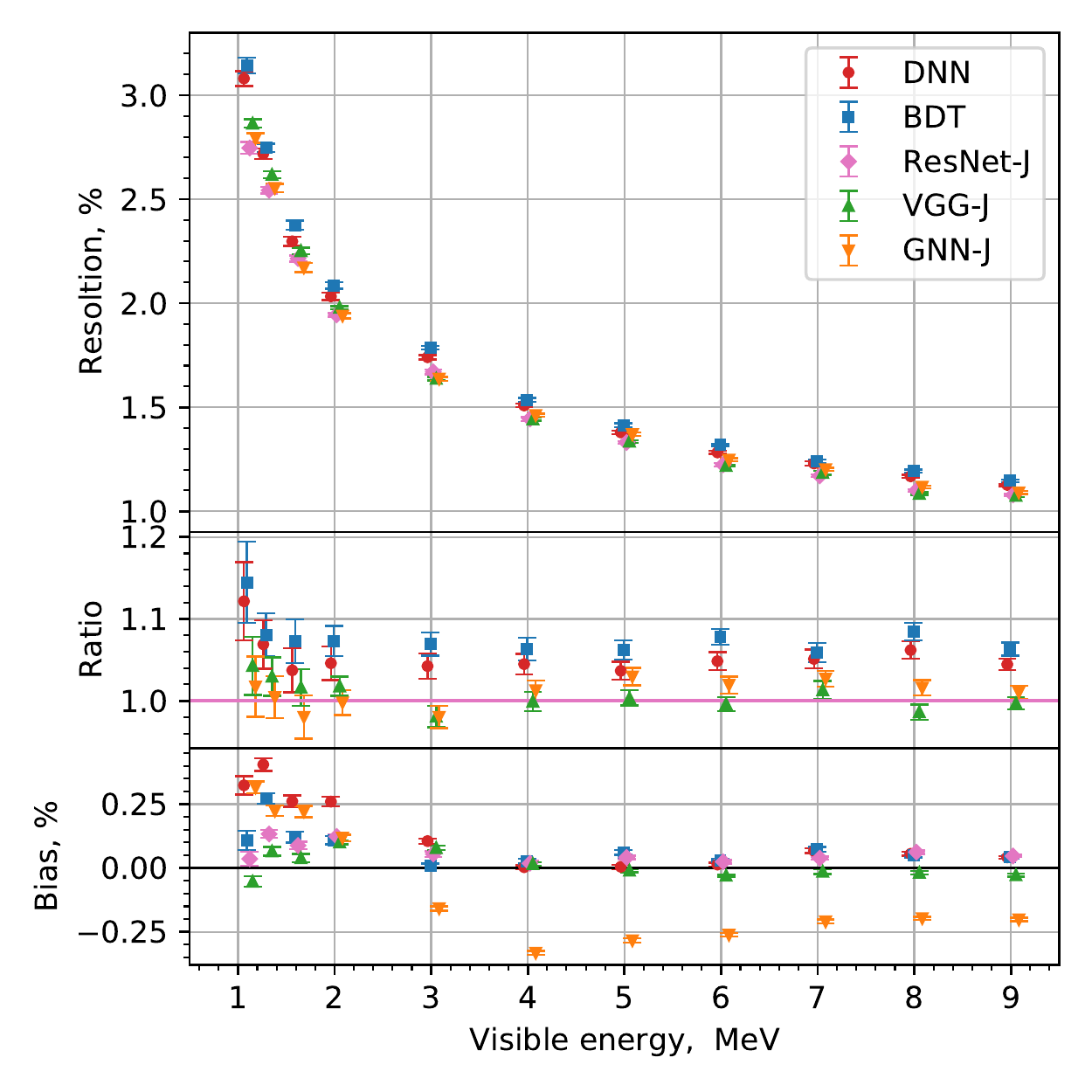}
    \caption{Energy reconstruction performance: resolution (upper panel) and bias (lower panel)
    obtained with DNN, BDT, ResNet-J, VGG-J and GNN-J model taking TTS and DN into account.
    The plots are offset along X-axis within $\pm\SI{0.06}{\MeV}$ for better readability. Note 
    the first point correspond to \SI{1.1}{MeV}.}
    \label{fig:ene_options}
\end{figure}

The performance comparison of BDT, DNN, ResNet-J, VGG-J and GNN-J is shown in Figure~\ref{fig:ene_options}
for the realistic case that takes TTS and DN into account. In general the complex models (ResNet-J, VGG-J 
and GNN-J) perform better. Their resolution is systematically finer than the one achieved with DNN and BDT.
ResNet-J and VGG-J also yield the lowest bias which lies within 0.15\%, while for the other models it is 
within 0.4\%. 

Interestingly BDT and DNN, using much less input, exhibit almost the same resolution as the complex models.
However, even such a small gain obtained with CNNs and GNN is crucial for JUNO.

%

%% file: 04_comp_time.tex
\subsection{Computation performance}

\begin{figure}[htb]
\centering
    \centering
    \includegraphics[width=0.6\linewidth]{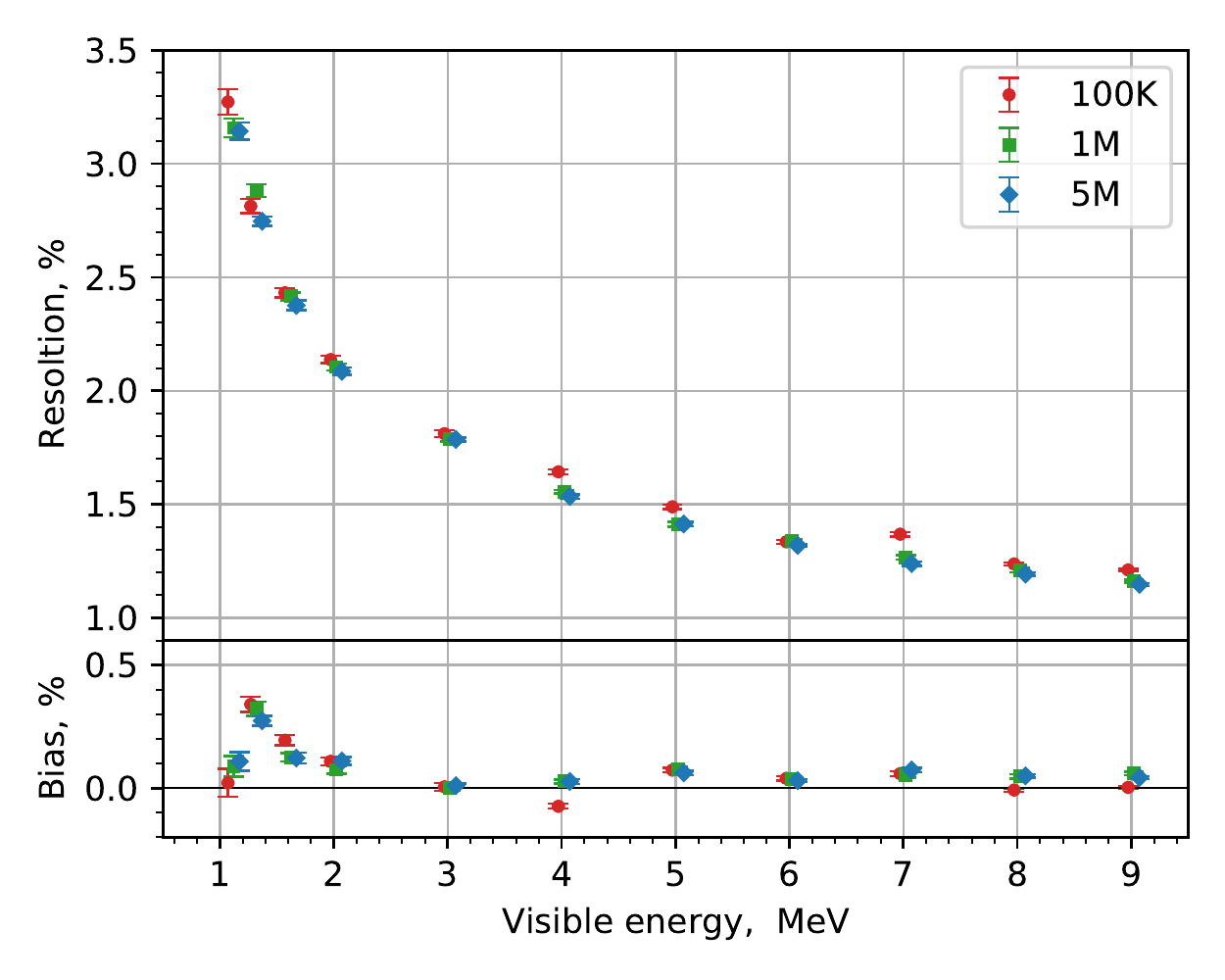}
    \caption{BDT performance for 100 thousand, 1 million and 5 million events in
        the training dataset. The plots are offset along X-axis within $\pm\SI{0.06}{\MeV}$ for better readability.}
    \label{fig:bdt_stat}
\end{figure}

BDT are known for their ability to train fast in terms of number of events.
We have investigated how the accuracy degrades with lowering of the dataset size
from default 5~millions to 1~million and 100~thousands, see Figure~\ref{fig:bdt_stat}.
Although there is a difference between all the three options for the ideal case
(neglecting DN and TTS, not shown), for the most realistic case taking TTS and DN into account there is no
noticeable gain from using more than 1M events.
Table~\ref{tab:description} shows typical number of trees (estimators), training time, prediction time and memory usage for each BDT model trained on different amounts of events and used for the energy reconstruction.

\begin{table}[htb]
  \centering
  \scalebox{1}{
    \begin{tabular}{lrrr}
      \toprule
                           & 100k & 1M  & 5M  \\
                           \midrule
      Number of trees      & 150  & 150 & 600 \\
      Training time, min   & 1    & 5   & 90  \\
      Prediction time, sec & 1    & 1   & 5   \\
      Memory usage, MB     & 4    & 17  & 60  \\
      \bottomrule
  \end{tabular}}
  \caption{Typical training and prediction time, number of trees, memory usage for the BDT model.}
  \label{tab:description}
\end{table}

Table~\ref{tab:comp_time} shows prediction and training time and memory usage for all the models
studied in this work. All the models except BDT were trained and run to yield predictions on
NVidia V100 graphical card. BDT, due to its minimalistic nature, was run on a laptop with
Intel i5-7300HQ CPU (2.5 GHz).
Since the batch size affects the prediction time and is individually optimized
for each model it is also presented in the table.
Training time is highly affected by the choice of the stopping rule. It was not common
for our models, so the numbers in the table are listed only to give a rough idea and can not be
used for a direct comparison.

Nonetheless, general conclusions can be drawn. The simple models are much faster and are advised
to be used for the tasks not requiring the finest possible resolution. BDT is especially fast.
The complex models require much more computational resources, GNN-J being somewhat less demanding
than ResNet-J and VGG-J.


\begin{table}[!htb]
    \small
    \centering
    \renewcommand{\arraystretch}{1.2}
    \begin{tabular}{@{}m{59mm}@{\hspace{5mm}}rrrrr@{}}
      \toprule
                                              &                         &              & \multicolumn{2}{c}{Planar CNN} & Spherical      \\
        \cmidrule{4-5}
        Architecture                          & \multicolumn{1}{r}{BDT} & DNN          & ResNet-J                       & VGG-J           & GNN-J        \\
        \midrule
        Prediction time, sec/100k events      & $<$1                    & $<$1         & 235                            & 155             & 110          \\
        Prediction batch size                 & \num{100000}            & \num{100000} & 100                            & 100             & \num{10000}  \\
        Number of weights                     &                         & 6625         & \num{38352403}                 & \num{26310035}  & \num{353979} \\
        Memory occupied by weights, MB        & 17                      & 0.073        & 146                            & 100             & 4.2          \\
        Training time, min/1M events          & 5                       & 1000         & 1543                           & 840             & 265          \\
        Training batch size                   &                         & 700          & 64                             & 64              & 64           \\
        \bottomrule
    \end{tabular}
    \caption{Prediction time and memory usage for different models. BDT was run on an ordinary laptop,
while the other models were run on graphical card NVidia V100. The preparation of data
is computationally expensive, but since we did not undertake any efforts for its optimization in some cases,
it is not shown in the table.}
    \label{tab:comp_time}
\end{table}

%% file: 05_discussion.tex
\subsection{Fine-tuning with calibration data}

JUNO experiment will include a calibration campaign~\cite{JUNO_calib}: different
radio-active sources will be introduced into the detector volume at various positions
to study the detector response. A large calibration is planed to be performed before
starting physics data taking to perform the initial evaluation of the detector.
Shorter calibrations will be performed regularly to monitor changes in the detector response.
The obtained information will be used to improve the computer simulation of the detector. 


Although the JUNO simulation will be made as realistic as possible, there is a chance 
that the predictions of ML models trained exclusively on MC will be biased. 
In order to fine-tune ML models the calibration data may be used directly:
\begin{enumerate}
  \item Train ML models on a large amount of Monte-Carlo data.
  \item Check how biased are the predictions on the calibration data.
\end{enumerate}%
If the prediction quality is not acceptable:
\begin{enumerate}
\setcounter{enumi}{2}
\item Continue training on a half (or other fraction) of calibration data.
\item Verify its accuracy with the rest of calibration data.
\end{enumerate}

It was shown~\cite{tajbakhsh2016convolutional} that such an approach could yield a 
good performance even when trained on a small sample of data compared to the original 
Monte Carlo. The problem is that neither the energy range nor the detector
volume is fully covered by the calibration. In other words, the data
comes from a different domain, compared to the data that the ML model were trained on. 
In case fine-tuning fails, there is a possibility to implement domain adaptation techniques~\cite{ganin2015unsupervised,tzeng2017adversarial} to overcome
the domain shift and dataset bias problems.




\subsection{Prospects}

We see several ways that could be used in order to improve the precision and the prediction speed of the models. 

The further research includes: 
adding extra feature
variables as new input to DNN and BDT methods; reducing the number of layers and parameters of VGG-J and ResNet-J to trade-off between better
reconstruction precision and fewer computing resources. 

Another interesting possibility is to generalize the GNN-J model so that 
no spherical discretization is used, by using a graph coarsening
algorithm such as Graclus~\cite{graclus}. This would enable us to use the data 
from PMTs directly without any need of
aggregation to an initial resolution, as it is done in the current implementation. 

Currently, only the time of the first hit is used. In~\cite{allhittimerec} it was pointed out that, taking into account the time of all hits should improve the precision of vertex reconstruction. 
Thus, the precision of the waveform
reconstruction providing charge and time information is highly important, especially when it comes to the reconstruction
of the time of all the hits in the event. The application of machine learning techniques for the waveform
reconstruction looks promising in terms of precision and numerical performance.

To improve the performance of the energy reconstruction, extra information provided by \num{25600} $3''$ PMTs in the CD can
be used. Special methods, both classical and ML, may be used to remove the DN counts from the signal which may
further reduce the energy reconstruction uncertainty.


%% file: 06_conclusions.tex
In this work we present several novel strategies of the event vertex and energy reconstruction 
in JUNO detector using several machine learning methods trained on Monte Carlo simulation 
data. We have studied the following approaches: boosted decision trees (BDT), deep neural 
networks (DNN) and a few kinds of convolution neural networks (CNN) including graph 
neural network (GNN). 
After proper tuning we could achieve vertex and energy resolution satisfying the requirements
posed by the physical goals of the experiment.

The study takes into account the electronics effects: the dark noise and the transition time spread.
The former one is found to be the main effect worsening the performance of
the energy reconstruction while the latter one has the main effect on the vertex reconstruction.

For the vertex reconstruction it was found to be crucial to provide PMT-wise 
information to the model:
the planar CNN models (ResNet-J and VGG-J) have 4 times better performance than minimalistic
models (BDT and DNN) working with several aggregated features. The best achieved resolution 
of the vertex coordinates is around $\sigma_{x,y,z}=10$~cm at $E_{\rm vis}=1$~MeV
and decreases at higher energies.

The minimalistic models (BDT and DNN), two implementations of CNNs (ResNet-J and VGG-J) and the 
spherical model GNN-J provide the energy resolution of around $\sigma_E=3\%$ at $E_\text{vis}=1$~MeV.
However, the latter three show slightly better performance thanks to PMT-wise input information.

We plan to continue improving the models: searching for new informative aggregated features for
DNN and BDT, optimizing the structure of ResNet-J and VGG-J, and developing the no-projection 
version of GNN-J.

Future work will include the investigation of the applicability of the models trained on the simulation
data to the real data collected by the JUNO detector. Methods for the inclusion of the calibration
data will be studied.

%% file: 07_acknowledgement.tex
We would like to thank Weidong Li, Jiaheng Zou, Tao Lin, Ziyan Deng, Guofu Cao
and Miao Yu for their tremendous contribution to the development of JUNO
offline software and to Xiaomei Zhang and Jo\~ao Pedro Athayde Marcondes de
Andr\'e for the production of the MC samples. 

We are grateful to N.~Kutovskiy, N.~Balashov for providing an extensive IT
support and computing resources of JINR cloud services~\cite{Baranov:2016gvt}
and to D.~Podgainy, O.~Streltsova, D.~Belyakov, M.~Matveev for providing the
dedicated GPU-enabled queue on HybriLIT platform
(LIT,~JINR)~\cite{HybriLIT:2018} for this work. CloudVeneto is acknowledged for
the use of computing and storage facilities.
This research was supported in part through computational resources of HPC facilities at NRU HSE.
We also deeply appreciate the help from the Computing Center of the Institute
of High Energy Physics, Chinese Academy of Science for providing the GPU
resources. 

This work is supported by the National Natural Science Foundation of China (No. 11805294, 11975021),
the China Postdoctoral Science Foundation (2018M631013),
the Strategic Priority Research Program of Chinese Academy of Sciences (XDA10010900),
the Fundamental Research Funds for the Central Universities, Sun Yat-sen University (19lgpy268).
Maxim Gonchar and Yury Malyshkin are supported by the Ministry of science and higher education 
of the Russian Federation under the contract \textnumero{}075-15-2020-77.  
Vladislav Belavin, Andrey Ustyuzhanin, and Fedor Ratnikov are supported by the
Russian Science Foundation under grant agreement \textnumero{}17-72-20127 for their work on DNN methods and their optimization.
Konstantin Treskov is supported by the grant \textnumero{}21-202-10 of JINR Association of Young Scientists and Specialists.